\documentclass[iop,apj,numberedappendix,appendixfloats]{emulateapj}

\newcommand{\noprint}[1]{}
\newcommand{\figsetstart}{{\bf Fig. Set} }
\newcommand{\figsetend}{}
\newcommand{\figsetgrpstart}{}
\newcommand{\figsetgrpend}{}
\newcommand{\figsetnum}[1]{{\bf #1.}}
\newcommand{\figsettitle}[1]{ {\bf #1} }
\newcommand{\figsetgrpnum}[1]{\noprint{#1}}
\newcommand{\figsetgrptitle}[1]{\noprint{#1}}
\newcommand{\figsetplot}[1]{\noprint{#1}}
\newcommand{\figsetgrpnote}[1]{\noprint{#1}}

\usepackage{apjfonts}
\usepackage{natbib}
\usepackage{amssymb}
\usepackage{enumitem}
\usepackage{comment}
\usepackage{multirow}
\usepackage[title]{appendix}
\usepackage[breaklinks,colorlinks,citecolor=blue,linkcolor=magenta]{hyperref}
\usepackage{color,soul}
\newcommand{\OVIdblt}{{\rm O}\kern 0.1em{\sc vi}~$\lambda\lambda 1031, 1037$} 
\newcommand{\MgIIdblt}{{\rm Mg}\kern 0.1em{\sc ii}~$\lambda\lambda 2796, 2803$}
\newcommand{\OVI}{\hbox{{\rm O}\kern 0.1em{\sc vi}}}
\newcommand{\OVII}{\hbox{{\rm O}\kern 0.1em{\sc vii}}}
\newcommand{\OVIII}{\hbox{{\rm O}\kern 0.1em{\sc viii}}}
\newcommand{\MgII}{\hbox{{\rm Mg}\kern 0.1em{\sc ii}}}
\newcommand{\CIV}{\hbox{{\rm C}\kern 0.1em{\sc iv}}}
\newcommand{\HI}{\hbox{{\rm H}\kern 0.1em{\sc i}}}
\newcommand{\PVdblt}{{\rm P}\kern 0.1em{\sc v}~$\lambda\lambda 1117, 1128$}
\newcommand{\CaIIdblt}{{\rm Ca}\kern 0.1em{\sc ii}~$\lambda\lambda 3934, 3969$}
\newcommand{\AlIIIdblt}{{\rm Al}\kern 0.1em{\sc iv}~$\lambda\lambda 1855, 1863$}
\newcommand{\CIVdblt}{{\rm C}\kern 0.1em{\sc iv}~$\lambda\lambda 1548, 1550$}
\newcommand{\NVdblt}{{\rm N}\kern 0.1em{\sc v}~$\lambda\lambda 1238, 1242$}  
\newcommand{\SVIdblt}{{\rm S}\kern 0.1em{\sc vi}~$\lambda\lambda 933, 944$} 
\newcommand{\SiIIdblt}{{\rm Si}\kern 0.1em{\sc ii}~$\lambda\lambda $1190$~{\AA}, $1193$~{\AA}$} 
\newcommand{\SiIVdblt}{{\rm Si}\kern 0.1em{\sc iv}~$\lambda\lambda 1393, 1402$} 
\newcommand{\PV}{\hbox{{\rm P}\kern 0.1em{\sc v}}}
\newcommand{\AlI}{\hbox{{\rm Al}\kern 0.1em{\sc i}}}
\newcommand{\AlII}{\hbox{{\rm Al}\kern 0.1em{\sc ii}}}
\newcommand{\AlIII}{{\hbox{\rm Al}\kern 0.1em{\sc iii}}}
\newcommand{\CaII}{\hbox{{\rm Ca}\kern 0.1em{\sc ii}}}
\newcommand{\CII}{\hbox{{\rm C}\kern 0.1em{\sc ii}}}
\newcommand{\CIIe}{\hbox{{\rm C$^{\ast}$}\kern 0.1em{\sc ii}}}
\newcommand{\CIII}{\hbox{{\rm C}\kern 0.1em{\sc iii}}}
\newcommand{\CV}{\hbox{{\rm C}\kern 0.1em{\sc v}}}
\newcommand{\HII}{\hbox{{\rm H}\kern 0.1em{\sc ii}}}
\newcommand{\Lya}{\hbox{{\rm Ly}\kern 0.1em$\alpha$}}
\newcommand{\Lyb}{\hbox{{\rm Ly}\kern 0.1em$\beta$}}
\newcommand{\Lyg}{\hbox{{\rm Ly}\kern 0.1em$\gamma$}}
\newcommand{\Lyd}{\hbox{{\rm Ly}\kern 0.1em$\delta$}}
\newcommand{\Lye}{\hbox{{\rm Ly}\kern 0.1em$\epsilon$}}
\newcommand{\Lyphi}{\hbox{{\rm Ly}\kern 0.1em$\phi$}}
\newcommand{\Lyfive}{\hbox{{\rm Ly}\kern 0.1em$5$}}
\newcommand{\Lysix}{\hbox{{\rm Ly}\kern 0.1em$6$}}
\newcommand{\Lyseven}{\hbox{{\rm Ly}\kern 0.1em$7$}}
\newcommand{\Lyeight}{\hbox{{\rm Ly}\kern 0.1em$8$}}
\newcommand{\Lynine}{\hbox{{\rm Ly}\kern 0.1em$9$}}
\newcommand{\Lyten}{\hbox{{\rm Ly}\kern 0.1em$10$}}
\newcommand{\Lyeleven}{\hbox{{\rm Ly}\kern 0.1em$11$}}
\newcommand{\HeI}{\hbox{{\rm He}\kern 0.1em{\sc i}}}
\newcommand{\HeII}{\hbox{{\rm He}\kern 0.1em{\sc ii}}}
\newcommand{\FeI}{\hbox{{\rm Fe}\kern 0.1em{\sc i}}}
\newcommand{\FeII}{\hbox{{\rm Fe}\kern 0.1em{\sc ii}}}
\newcommand{\FeIII}{\hbox{{\rm Fe}\kern 0.1em{\sc iii}}}
\newcommand{\MnII}{\hbox{{\rm Mn}\kern 0.1em{\sc ii}}}
\newcommand{\MgI}{\hbox{{\rm Mg}\kern 0.1em{\sc i}}}
\newcommand{\MgIII}{\hbox{{\rm Mg}\kern 0.1em{\sc iii}}}
\newcommand{\NI}{\hbox{{\rm N}\kern 0.1em{\sc i}}}
\newcommand{\NII}{\hbox{{\rm N}\kern 0.1em{\sc ii}}}
\newcommand{\NIII}{\hbox{{\rm N}\kern 0.1em{\sc iii}}}
\newcommand{\NV}{\hbox{{\rm N}\kern 0.1em{\sc v}}}
\newcommand{\OI}{\hbox{{\rm O}\kern 0.1em{\sc i}}}
\newcommand{\OII}{\hbox{[{\rm O}\kern 0.1em{\sc ii}]}}
\newcommand{\OIV}{\hbox{{\rm O}\kern 0.1em{\sc iv}]}}
\newcommand{\SI}{{\rm S}\kern 0.1em{\sc i}}
\newcommand{\SIV}{{\rm S}\kern 0.1em{\sc iv}}
\newcommand{\SVI}{{\rm S}\kern 0.1em{\sc vi}}
\newcommand{\SiI}{\hbox{{\rm Si}\kern 0.1em{\sc i}}}
\newcommand{\SiII}{\hbox{{\rm Si}\kern 0.1em{\sc ii}}}
\newcommand{\SiIII}{\hbox{{\rm Si}\kern 0.1em{\sc iii}}}
\newcommand{\SiIV}{\hbox{{\rm Si}\kern 0.1em{\sc iv}}}
\newcommand{\SII}{\hbox{{\rm S}\kern 0.1em{\sc ii}}}
\newcommand{\SIII}{\hbox{{\rm S}\kern 0.1em{\sc iii}}}
\newcommand{\NaI}{\hbox{{\rm Na}\kern 0.1em{\sc i}}}
\newcommand{\TiII}{\hbox{{\rm Ti}\kern 0.1em{\sc ii}}}
\newcommand{\cms}{\hbox{cm$^{-2}$}}
\newcommand{\kms}{\hbox{km~s$^{-1}$}}
\newcommand{\etal}{et~al.}
\newcommand{\NHI}{\log N_{\hbox{\tiny \HI}}}

\shorttitle{CGM Metallicities in Group Environments}
\shortauthors{\sc Pointon {\etal}}

\begin{document}

\title{Low Mass Group Environments have no Substantial Impact on the Circumgalactic Medium Metallicity}
\author{
Stephanie K. Pointon$^{1,2}$
}
\author{Glenn G. Kacprzak$^{1,2}$
}
\author{
Nikole M. Nielsen$^{1,2}$
}
\author{
Michael T. Murphy$^{1}$
}
\author{Sowgat Muzahid$^{3}$
}
\author{Christopher W. Churchill$^{4}$
}
\author{Jane C. Charlton$^{5}$
}

\affil{$^1$ Centre for Astrophysics and Supercomputing, Swinburne
  University of Technology, Hawthorn, Victoria 3122, Australia;
  spointon@swin.edu.au
}
\affil{$^2$ ARC Centre of Excellence for All Sky Astrophysics in 3 Dimensions (ASTRO 3D)
}
\affil{$^3$ Leiden Observatory, University of Leiden, PO Box 9513, NL-2300 RA Leiden, The Netherlands
}
\affil{
	$^4$ Department of Astronomy, New Mexico State University, Las Cruces, NM 88003, USA
}
\affil{$^5$ Department of Astronomy and Astrophysics, The Pennsylvania State University, State College, PA 16801, USA
}

\begin{abstract}
We explore how environment affects the metallicity of the circumgalactic medium (CGM) using 13 low mass galaxy groups (2--5 galaxies) at $\langle z_{abs}\rangle=0.25$ identified near background quasars. Using quasar spectra from HST/COS and from Keck/HIRES or VLT/UVES we measure column densities of, or determine limits on, CGM absorption lines. We use a Markov chain Monte Carlo approach with Cloudy to estimate metallicities of cool ($T\sim10^4$K) CGM gas within groups and compare them to CGM metallicities of 47 isolated galaxies. Both group and isolated CGM metallicities span a wide range ($-2<$[Si/H]$<0$), where the mean group ($-0.54\pm0.22$) and isolated ($-0.77\pm0.14$) CGM metallicities are similar. Group and isolated environments have similar distributions of {\HI} column densities as a function of impact parameter. However, contrary to isolated galaxies, we do not find an anti-correlation between {\HI} column density and the nearest group galaxy impact parameter. We additionally divided the groups by member luminosity ratios (i.e., galaxy--galaxy and galaxy--dwarf groups). While there was no significant difference in their mean metallicities, a modest increase in sample size should allow one to statistically identify a higher CGM metallicity in galaxy--dwarf groups compared to galaxy--galaxy groups. We conclude that either environmental effects have not played an important role in the metallicity of the CGM at this stage and expect that this may only occur when galaxies are strongly interacting or merging, or that some isolated galaxies have higher CGM metallicities due to past interactions. Thus, environment does not seem to be the cause of the CGM metallicity bimodality. 

\end{abstract}


\section{Introduction}
\label{sec:intro}
The gas surrounding galaxies outside their disks/interstellar medium (ISM) and residing within their virial radii is known as the circumgalactic medium \citep{tumlinson17}. Our understanding of the CGM has mainly been derived from studies of isolated galaxies revealing that within $1~\text{R}_{vir}$, the CGM contains a mass comparable to the ISM and is comprised of accreting, outflowing, and recycling gas \citep[e.g.][]{kacprzak08, kacprzak11kin, kacprzak16, chen10a, rudie12, thom11, tumlinson11, magiicat2, magiicat1, werk13, peeples14}.

It is expected that the CGM in group environments would be affected by galaxy--galaxy interactions, and hence, be more complex. The effects of galaxy--galaxy interactions are clearly visible as tidal streams in {\HI} emission around the M81/M82 galaxy group \citep{yun94, chynoweth08, blok18}. Further observations of {\HI} gas in the CGM have found evidence for interactions in the form of tidal streams, warped disks and high-velocity clouds \citep[e.g.][]{puche92, swaters97, rand00,fraternali02, chynoweth08, sancisi08, mihos12, wolfe13}. Additionally, absorption studies of group environment CGM gas have detected the presence of tidal streams or intragroup gas \citep{whiting06, kacprzak10b, nestor11, gauthier13, bielby17, peroux17, pointon17, nielsen18, chen19}. The tidal streams and increased star-formation rates that occur during mergers have been suggested to increase the halo gas mass and cross-section \citep{york86, rubin10}. Furthermore, FIRE simulations have demonstrated that intergalactic transfer is the dominant mode of gas accretion for $z<1$ \citep{angles16}. These combined results suggest that group environments cause the CGM to be disrupted, similar to the stellar components of interacting galaxies. Given the large inferred size of the CGM \citep[$\sim 200$~{kpc} for $L_{\ast}$ galaxies at redshifts $z<1.0$;][]{tumlinson11, werk14}, it is possible that the CGM will be influenced by a merger before the visible components of the host galaxy \citep{nielsen18}.

Using {\MgII} as a tracer of cool gas in cluster environments, \citet{lopez08} detected an overabundance of strong {\MgII} absorbers near clusters compared to field galaxies. A similar enhancement of weak {\MgII} absorbers beyond the cluster center was not observed, consistent with expectations that these absorbers should be destroyed by the hot cluster environment. The distributions of the weak and strong {\MgII} absorbers within the cluster is then evidence for a truncated cold gas halo, consistent with simulations \citep{padilla09, andrews13}.  

\cite{chen10a} investigated group environments using {\MgII} as a tracer of cool gas. In seven out of eight of the  group environments identified, {\MgII} was detected. While the group environment {\MgII} absorption appeared to span the same equivalent width versus impact parameter range as isolated galaxies, the authors did not detect a significant anti-correlation. This is contrary to the strong and well-known anti-correlation for isolated galaxies \citep[e.g.][]{lanzetta90, steidel94, kacprzak08, kacprzak12b, chen08, chen10b, bordoloi11, magiicat2}. Indeed, \citet{magiicat2} found the anti-correlation between {\MgII} equivalent width and impact parameter for isolated galaxies to be highly significant ($7.9\sigma$).

Further studies have found that the radial distribution of {\MgII} is flatter in group environments compared to isolated galaxies \citep[e.g.][]{bordoloi11, nielsen18}. \citet{bordoloi11} found that the average {\MgII} equivalent widths decreased beyond $140$~{kpc} in group environments, whereas they began to decrease beyond $70$~{kpc} for isolated galaxies. They further found that the radial distribution for the group environments CGM is consistent with a superposition of individual overlapping halos. Thus the authors suggested that the group environment CGM is not strongly influenced by tidal stripping or outflows driven by increased star-formation. However, using the kinematic structure of {\MgII} absorbers in group environments, \citet{nielsen18} found that a superposition model can reproduce the equivalent widths required, but over-predicts absorption at high velocities due to the large velocity separations between the galaxies in the group. Instead, the authors suggest that the cool gas in group environments forms an intragroup medium, created by intergalactic transfer or tidal stripping. 

Major mergers are able to disrupt the structure of involved galaxies more than minor mergers. Thus is it possible that the type of merger/interaction affects the CGM gas differently. \citet{nielsen18} found that galaxy--galaxy groups (where the two brightest galaxies have similar luminosities, $L_1/L_2 < 3.5$) may have larger equivalent widths ($1.7\sigma$) and absorber velocity dispersions ($2.5\sigma$) than galaxy--dwarf groups ($L_1/L_2 \geq 3.5$), while the covering fractions for the two samples are consistent within uncertainties. They suggest that tidal stripping of CGM gas and increased star formation might be more likely to occur in galaxy--galaxy groups.

The cool gas in the CGM, traced by {\MgII}, is likely to be constrained to high density structures surrounded by highly ionized gas, traced by {\CIV} and {\OVI}. This highly ionized gas has also been investigated in group environments \citep[e.g.][]{stocke13,burchett16, pointon17, ng19}. \citet{burchett16} found that as the number of galaxies in a group increases, the {\CIV} equivalent width decreases, with no {\CIV} detected in groups with more than seven galaxies. Similarly, {\OVI} has lower velocity spreads and column densities in group environments compared to isolated environments \citep{stocke13, pointon17, ng19}. These results are consistent with the picture that the virial temperature, which scales with halo mass, leads to oxygen and carbon ionising to higher states than {\OVI} and {\CIV}, respectively \citep{oppenheimer16, bielby18, zahedy18, ng19}.

All of this evidence suggests that it is possible for CGM metallicities to also be impacted by environment. Simulations by \citet{hani18} investigated the changes in CGM metallicity during a major merger. The authors found that, compared to the pre-merger state, the metallicity of the gas increased during the merger by $0.2-0.3$~dex. The increase was driven by outflows from the increased star-formation caused by the merger, rather than tidal stripping, and this metallicity level was maintained for several billions of years post-merger. This evidence that major mergers are capable of changing the CGM metallicity provides incentive for studying the metallicity of group environments pre-merger as a baseline.

Preliminary results from \citet{lehner18review} compare high and low metallicity absorbers from both isolated and group environments. They found that for partial Lyman limit systems (pLLs; $16.2~{\cms} < {\NHI} < 17.2~{\cms}$) and Lyman limit systems (LLS; $17.2~{\cms} < {\NHI} < 19.0~{\cms}$), the high metallicity systems are more likely to be associated with group environments while the low metallicity systems are associated with isolated environments. While the authors cautioned that this result is preliminary and refrained from making any interpretations, it may suggest that interactions in groups of galaxies may be causing increased metallicity. This result is somewhat challenged by \citet{pointon19}, who studied the metallicity of the CGM in isolated environments. They found that the CGM metallicities of isolated galaxies span the full range detected by \citet{lehner18review}, even when the sample is restricted to the same {\HI} column density range. This suggests that high metallicity systems are not only found in group environments.

Following on from \citet{lehner18review}, we investigate the effect of environment on the metallicity of the CGM by comparing the isolated galaxy sample from \citet{pointon19} to group environments. We investigate the metallicity of 13 group environments using the combination of UV spectra from \textit{HST}/COS and \textit{FUSE}, as well as optical spectra from Keck/HIRES and VLT/UVES. 

\begin{deluxetable*}{lcllccccc}
	\tablecolumns{9}
	\tablewidth{0pt}
	\setlength{\tabcolsep}{0.06in}
	\tablecaption{Quasar Observations \label{tab:obsqso}}
	\tablehead{
		\colhead{(1)}           	&
        \colhead{(2)}     &
		\colhead{(3)}        &
		\colhead{(4)}       &
		\colhead{(5)}	&
		\colhead{(6)}	 	&				
		\colhead{(7)} 	&		
		\colhead{(8)} 	&
        \colhead{(9)}\\
		\colhead{J-Name}           	&
        \colhead{$z_{\rm qso}$}     &
		\colhead{RA (J2000)}        &
		\colhead{DEC (J2000)}       &
		\colhead{UV Inst.}	&
		\colhead{COS Gratings}	&
		\colhead{COS PID(s)}	 	&			
		\colhead{Optical Spectrograph} 	&
        \colhead{Optical PID(s)}}
	\startdata
	J$0125$	&	$1.074$	&	$01$:$25$:$28.84$	&	$-00$:$05$:$55.93$	&   COS &	G160M	        &	13398	&	UVES	    &	075.A-0841(A)	\\
J$0228$ &   $0.493$ &   $02$:$28$:$15.17$   &   $-40$:$57$:$14.29$  &   COS &   G130M, G160M    &   11541   &   $\cdots$    &   $\cdots$        \\
J$0351$	&	$0.616$	&	$03$:$51$:$28.54$	&	$-14$:$29$:$08.71$ 	&   COS &	G130M, G160M	&	13398	&	UVES	    &	076.A-0860(A)	\\
J$0407$	&	$0.572$	&	$04$:$07$:$48.43$	&	$-12$:$11$:$36.66$	&   COS, FUSE &	G130M, G160M	&	11541	&	HIRES	    &	G01H, U68H	    \\
J$0853$	&	$0.514$	&	$08$:$53$:$34.25$	&	$+43$:$49$:$02.33$	&   COS &	G130M, G160M	&	13398	&	$\cdots$	&	$\cdots$	    \\
J$0910$ &   $0.463$ &   $09$:$10$:$29.75$   &   $+10$:$14$:$13.61$  &   COS &   G130M, G160M    &   11598   &   $\cdots$	&	$\cdots$	    \\
J$0925$ &   $0.472$ &   $09$:$25$:$54.71$   &   $+40$:$04$:$14.17$  &   COS &   G130M, G160M    &   11598   &   HIRES         &	U059Hb	            \\
J$0928$ &   $0.296$ &   $09$:$28$:$37.98$   &   $+60$:$25$:$21.02$  &   COS &   G130M, G160M    &   11598   &    HIRES        &   U066Hb                  \\
J$1009$	&	$0.456$	&	$10$:$09$:$02.06$	&	$+07$:$13$:$43.87$	&   COS &	G130M, G160M	&	11598	&	HIRES	    &	U066Hb	        \\
J$1119$	&	$0.176$	&	$11$:$19$:$08.67$	&	$+21$:$19$:$18.01$	&   COS, FUSE &	G130M, G160M	&	12038	&	HIRES	    &	U152Hb	        \\
J$1139$	&	$0.556$	&	$11$:$39$:$10.70$	&	$-13$:$50$:$43.63$	&   COS &	G130M	        &	12275	&	$\cdots$	&	$\cdots$	    \\[-10pt]

	\enddata
\end{deluxetable*}

This paper is organized as follows: In Section \ref{sec:observations} we describe our sample of group galaxy--absorber pairs. We also describe how we obtain the metallicity of the CGM. We present the results comparing the group environment CGM metallicity with the same properties for isolated galaxies, as well as investigate any trends with {\HI} column density, impact parameter and luminosity in Section \ref{sec:results} and discuss the implications in Section \ref{sec:discussion} . In Section \ref{sec:conclusions} we summarize our results and provide concluding remarks. We use a standard $\Lambda$CDM cosmology with $H_o = 70$~{\kms} Mpc$^{-1}$, {$\Omega_M = 0.3$} and {$\Omega_\Lambda = 0.7$}.

\section{Observations}\label{sec:observations}
  \begin{deluxetable*}{llcrrrrrrr}
      \tablecolumns{10}
      \tablecaption{Galaxy Properties \label{tab:obsgal}}
      \tablehead{
          \colhead{(1)}     &
          \colhead{(2)}     &
          \colhead{(3)}     &
          \colhead{(4)}     &
          \colhead{(5)}     &
          \colhead{(6)}	    &
          \colhead{(7)}     &
          \colhead{(8)}     &
          \colhead{(9)}     &    
          \colhead{(10)}\\
          \colhead{J-Name}            &
          \colhead{$z_{\rm gal}$}     &
          \colhead{REF\tablenotemark{a}}               &
          \colhead{$\Delta \alpha$ }  &
          \colhead{$\Delta \delta$}   &
          \colhead{$\theta$}          &
          \colhead{D}	              &
          \colhead{$M_{B}$}          &
          \colhead{$L_B/L_{B}^{*}$}  &
          \colhead{$v_{G1}-v_{GX}$\tablenotemark{b}}\\
          \colhead{}            &
          \colhead{}            &
          \colhead{}            &
          \colhead{(J2000)}     &
          \colhead{(J2000)}     &
          \colhead{(deg)}       &
          \colhead{(kpc)}	    &
          \colhead{}            &
          \colhead{}	        &
          \colhead{(\kms)}
          }
      \startdata
      \rule{0pt}{3ex}\multirow{2}{*}{J0125}	&$	0.3787	$&$ (1) $&$	-8.8	$&$	-12.3	$&$	15.07	$&$	78	 $& $-20.21 $ & $0.52 $&$ - 	$\\
					                    &$	0.3792	$&$ (2) $&$	-27.7	$&$	-36.5	$&$	45.80	$&$	238	 $& $-20.21 $ & $0.57 $&$ 108	$\\\hline
\rule{0pt}{3ex}\multirow{2}{*}{J0228}	&$	0.2065	$&$ (3) $&$	-9.1	$&$	-8.4	$&$	10.87	$&$	34	 $& $-19.42 $ & $0.32 $&$ - 	$\\
					                    &$	0.2078	$&$ (3) $&$	-24.9	$&$	-25.9	$&$	32.04	$&$	109	 $& $-18.32 $ & $0.15 $&$ 323	$\\\hline
\rule{0pt}{3ex}\multirow{3}{*}{J0228}	&$	0.2678	$&$ (3) $&$	16.9	$&$	-13.0	$&$	18.21	$&$	63	 $& $-19.43 $ & $0.29 $&$ - 	$\\
					                    &$	0.2690	$&$ (3) $&$	8.5	    $&$	-36.7	$&$	37.26	$&$	154	 $& $-16.78 $ & $0.02 $&$ 284	$\\
				                    	&$	0.2680	$&$ (3) $&$	36.2	$&$	-29.2	$&$	39.98	$&$	164	 $& $-18.01 $ & $0.08 $&$ 47	$\\\hline
\rule{0pt}{3ex}\multirow{3}{*}{J0351}	&$	0.324180$&$ (4) $&$	13.0	$&$	-23.5	$&$	26.72	$&$	126	 $& $-20.15 $ & $0.52 $&$ - 	$\\
					                    &$	0.324651$&$ (4) $&$	-29.9	$&$	18.5	$&$	34.33	$&$	162	 $& $-20.95 $ & $1.09 $&$ 107	$\\
					                    &$	0.3273	$&$ (2) $&$	-59.0	$&$	19.5	$&$	60.31	$&$	288	 $& $-20.27 $ & $0.58 $&$ 706	$\\\hline
\rule{0pt}{3ex}\multirow{5}{*}{J0407}	&$	0.0923	$&$ (5) $&$	13.6	$&$	-39.9	$&$	42.08	$&$	72	 $& $-15.45 $ & $0.01 $&$ - 	$\\
					                    &$	0.0908	$&$ (5) $&$	-61.9	$&$	13.6	$&$	62.01	$&$	105	 $& $-15.88 $ & $0.01 $&$ -412	$\\
					                    &$	0.0914	$&$ (5) $&$	-78.9	$&$	-10.6	$&$	77.89	$&$	133	 $& $-15.84 $ & $0.01 $&$ -247	$\\
					                    &$	0.0917	$&$ (5) $&$	-123.5	$&$	-127.3	$&$	175.44	$&$	300	 $& $-15.12 $ & $0.01 $&$ -165	$\\
					                    &$	0.0908	$&$ (5) $&$	62.5	$&$	-252.3	$&$	259.64	$&$	439	 $& $-17.29 $ & $0.05 $&$ -412	$\\\hline
\rule{0pt}{3ex}\multirow{2}{*}{J0407}	&$	0.16699 $&$ (4) $&$	-1.1	$&$	34.8	$&$	34.81	$&$	99	 $& $-18.04 $ & $0.09 $&$ - 	$\\
					                    &$	0.16699 $&$ (4) $&$	41.3	$&$	-1.8	$&$	40.36	$&$	115	 $& $-21.65 $ & $2.49 $&$ 0	$\\\hline
\rule{0pt}{3ex}\multirow{2}{*}{J0853}	&$	0.0903	$&$ (6) $&$	14.1	$&$	-34.0	$&$	35.45	$&$	79	 $& $-18.75 $ & $0.19 $&$ - 	$\\
					                    &$	0.0915	$&$ (3) $&$	-1.8	$&$	40.8	$&$	40.81	$&$	53	 $& $-17.28 $ & $0.05 $&$ 330	$\\\hline
\rule{0pt}{3ex}\multirow{2}{*}{J0910}	&$	0.2647	$&$ (7) $&$	8.3	    $&$	11.4	$&$	13.99	$&$	54	 $& $-19.70 $ & $0.37 $&$ - 	$\\
					                    &$	0.2641	$&$ (7) $&$	-30.8	$&$	-16.4	$&$	34.42	$&$	132	 $& $-21.00 $ & $1.23 $&$ -142	$\\\hline
\rule{0pt}{3ex}\multirow{2}{*}{J0925}	&$	0.2467	$&$ (7) $&$	-7.2	$&$	-24.1	$&$	24.69	$&$	96	 $& $-20.52 $ & $0.80 $&$ - 	$\\
				                    	&$	0.2475	$&$ (7) $&$	-8.0	$&$	-20.8	$&$	21.64	$&$	84	 $& $-21.25 $ & $1.57 $&$ 192	$\\\hline
\rule{0pt}{3ex}\multirow{3}{*}{J0928}	&$	0.1540	$&$ (7) $&$	67.2	$&$	-12.3	$&$	35.38	$&$	95	 $& $-20.14 $ & $0.63 $&$ - 	$\\
				                    	&$	0.1542	$&$ (7) $&$	30.2	$&$	-12.1	$&$	19.19	$&$	51	 $& $-19.84 $ & $0.48 $&$ 52	$\\
				                      	&$	0.1537	$&$ (7) $&$	-3.5	$&$	-14.7	$&$	14.82	$&$	40	 $& $-18.76 $ & $0.18 $&$ -78	$\\\hline
\rule{0pt}{3ex}\multirow{2}{*}{J1009}	&$	0.35587 $&$ (4) $&$	1.7	    $&$	-9.3	$&$	9.41	$&$	47	 $& $-19.98 $ & $0.43 $&$ - 	$\\
				                    	&$	0.35585 $&$ (4) $&$	3.2	    $&$	0.0	    $&$	3.13	$&$	16	 $& $-17.87 $ & $0.06 $&$ -4	$\\\hline
\rule{0pt}{3ex}\multirow{2}{*}{J1119}	&$	0.0600	$&$ (8) $&$	-48.1	$&$	-104.9	$&$	117.01	$&$	136	 $& $-17.75 $ & $0.08 $&$ - 	$\\
			                    		&$	0.0594	$&$ (8) $&$	-83.1	$&$	-174.6	$&$	190.99	$&$	219	 $& $-16.56 $ & $0.03 $&$ -170	$\\\hline
\rule{0pt}{3ex}\multirow{2}{*}{J1133}	&$	0.2367	$&$ (7) $&$	4.5	    $&$	-1.7	$&$	4.79	$&$	18	 $& $-21.24 $ & $1.58 $&$ - 	$\\
				                    	&$	0.2364	$&$ (7) $&$	-4.1	$&$	-9.6	$&$	10.39	$&$	39	 $& $-20.54 $ & $0.83 $&$ -73	$\\[-10pt]

      \enddata
      \tablenotetext{a}{Galaxy identification references: (1) \citet{muzahid15}, (2) \citet{chen01b}, (3) \citet{chen09}, (4) \citet{nielsen18}, (5) \citet{johnson15}, (6) \citet{lanzetta95}, (7) \citet{werk12} and (8) \citet{prochaska11}.}
      \tablenotetext{b}{Line-of-sight velocity separations between the first galaxy in the group ($G1$) and each of the other group galaxy members ($GX$).}
  \end{deluxetable*}

In order to study the CGM of the group environments, we use the ``Multiphase Galaxy Halos'' Survey which is comprised of UV \textit{HST}/COS spectra from our program (PID 13398) \citep{kacprzak15, kacprzak19, muzahid15, muzahid16, pointon17,  pointon19, nielsenovi,ng19} as well as data taken from literature \citep{chen01b, chen09, meiring11, werk12, johnson13}. A group environment is defined as having the nearest of two or more galaxies within $18$ to $150$~kpc of the quasar sight-line in order to replicate the impact parameter distribution of the isolated sample. The galaxies in the group must have line-of-sight velocity separations of less than $1000$~{\kms} and a maximum impact parameter of $500$~kpc. We investigate 13 group environments from the literature for which we have UV spectra \citep{lanzetta95, chen01b, chen09, prochaska11, werk12, johnson15, muzahid15, nielsen18}. The groups have associated {\HI} absorption with a redshift range of $0.06 < z_{abs} < 0.38$ ($\langle z_{abs} \rangle = 0.25$). The group environments have a wide range of luminosity ratios between the brightest ($L_1$) and second brightest ($L_2$) galaxies ($1.1 < L_1/L_2 < 27.7$; median $L_1/L_2 = 2.7$), indicating that we are investigating groups with a variety of mass ratios. Typical group environments in this study have two members, although J$0407$, $z_{abs} = 0.0914$ has five galaxies, with a mean of $2.2$ galaxies per group. We note that group environments range from galaxy--dwarf pairs up to clusters of galaxies. Therefore, our study probes the low mass end of group environments. 

All quasars in the sample have COS UV spectra, while two also have reduced UV spectra from the \textit{FUSE} telescope, provided by B. Wakker (2016, private communication). Eight quasars have optical spectra from Keck/HIRES or VLT/UVES. The details of the quasar spectra are shown in Table \ref{tab:obsqso}. 

\subsection{UV Quasar Spectra}\label{sec:cos}
The COS quasar spectra used in our survey have a median resolving power of $R \approx 20,000$, while the \textit{FUSE} quasar spectra have a resolving power of $R \approx 30,000$. The instruments, gratings and PID(s) for both COS and \textit{FUSE} quasar spectra are in Table \ref{tab:obsqso}. The range of ions covered by the UV spectra includes  the {\HI} Lyman series, {\CII}, {\CIII}, {\CIV}, {\NII}, {\NIII}, {\NV}, {\OI}, {\OVI}, {\SiII}, {\SiIII} and {\SiIV}. The reduction process for the \textit{HST}/COS spectra is described in detail in \citet{kacprzak15}. The raw data were reduced using the CALCOS pipeline software and then flux calibrated. Individual grating integrations were co-added and rebinned by three pixels to improve the signal-to-noise ratio \citep{danforth10}\footnote{\url{http://casa.colorado.edu/~danforth/science/cos/costools.html}}. The COS and \textit{FUSE} UV spectra were then continuum normalized by fitting low-order polynomials to the spectra while excluding absorption and emission lines from the fitting region. 

\subsection{Optical Quasar Spectra}
The UV spectra were complemented by additional optical spectra, which cover ionic transitions including {\MgI}, {\MgII}, {\FeII}, {\MnII} and {\CaII} for redshifts of $z > 0.2$. Eight quasars have optical spectra from Keck/HIRES and VLT/UVES with a resolving power of $R\approx 40,000$. The spectrograph and PID(s) for the optical spectra are in Table \ref{tab:obsqso}. IRAF or the Mauna Kea Echelle Extraction (MAKEE) package were used to reduce the HIRES data. The UVES spectra were reduced using the European Southern Observatory pipeline \citep{dekker00} and the UVES Post-Pipeline Echelle Reduction (UVES POPLER) code \citep{murphy16, murphy18}.

\subsection{Optical Galaxy Spectra}
Optical spectra of the galaxies in three group environment quasar fields were obtained using the Keck Echelle Spectrograph and Imager \citep[ESI;][]{sheinis02} since the wavelength range ($4000$--$10,000$\AA) provides coverage of emission lines including H$\alpha$. The reduction method is described in \citet{kacprzak19}, \citet{nielsen18} and \citet{pointon19}. However, we summarise the process here. The data, taken through slits of $20^{\prime\prime}$ by $1^{\prime\prime}$, were binned by two, resulting in a spatial pixel size of $0.27^{\prime\prime} - 0.34^{\prime\prime}$ and a spectral resolution of $22~\kms$. The reduction process was completed using IRAF, after which heliocentric and vacuum corrections were applied to the data.  Galaxy redshifts are shown in column (2) of Table \ref{tab:obsgal}. The redshifts for the remaining galaxies were obtained from literature, indicated in column (3) of Table \ref{tab:obsgal}.

\subsection{Isolated Galaxy Sample}
We use the metallicity study of 47 isolated environments by \citet{pointon19} with a redshift range of $0.06 < z < 0.66$ ($\langle z \rangle = 0.27$) to compare to the group environments. An isolated galaxy is defined as having no neighboring galaxies within a spatial separation of $150$~kpc and within a line-of-sight velocity separation of $1000$~{\kms}. Where the spatial or kinematic criteria were not met, the system was classified as a group environment. The impact parameters range from $18 < D < 203$~kpc. The isolated galaxies are roughly $L_{\ast}$ galaxies, with a halo mass range of $10.8 < \log M_{h}/M_{\odot} < 12.5$, ($\langle \log M_{h}/M_{\odot} \rangle = 11.8$). The absorption systems span {\HI} column densities ranging from $13.8 <\NHI<19.9$. Using the same methods we use here, \citet{pointon19} estimated CGM metallicities ranging from $-2.6 <$~[Si/H]~$<0.8$ with an average of $\langle$[Si/H]$\rangle = -1.3$.

\subsection{Sample Comparison}
The sample investigated here is a collation of quasar fields that have been previously spectroscopically surveyed \citep[see Table \ref{tab:obsgal} and][]{pointon19}. Consequently, each survey has different levels of completeness but typically have a luminosity sensitivity of $0.1L_{\ast}$. Fields drawn from the COS Halos survey have been probed out to a distance of $150$~kpc \citep[see][]{tumlinson13,werk13}, while other fields have been investigated out to at least $350$~kpc \citep[see][for further details]{pointon17, nielsen18, kacprzak19, ng19}. It is possible that isolated galaxies identified in the COS Halos survey may be a member of a group which extends beyond the survey regions. To investigate this, we repeated all statistical tests with the COS Halos galaxies removed from the isolated sample. We do not find any difference in results with the COS Halos fields removed and hence, included all galaxies in our full isolated sample (Tables \ref{tab:ADtest} and \ref{tab:KTtest}).

\begin{figure*}
	\centering
	\includegraphics[width=\linewidth]{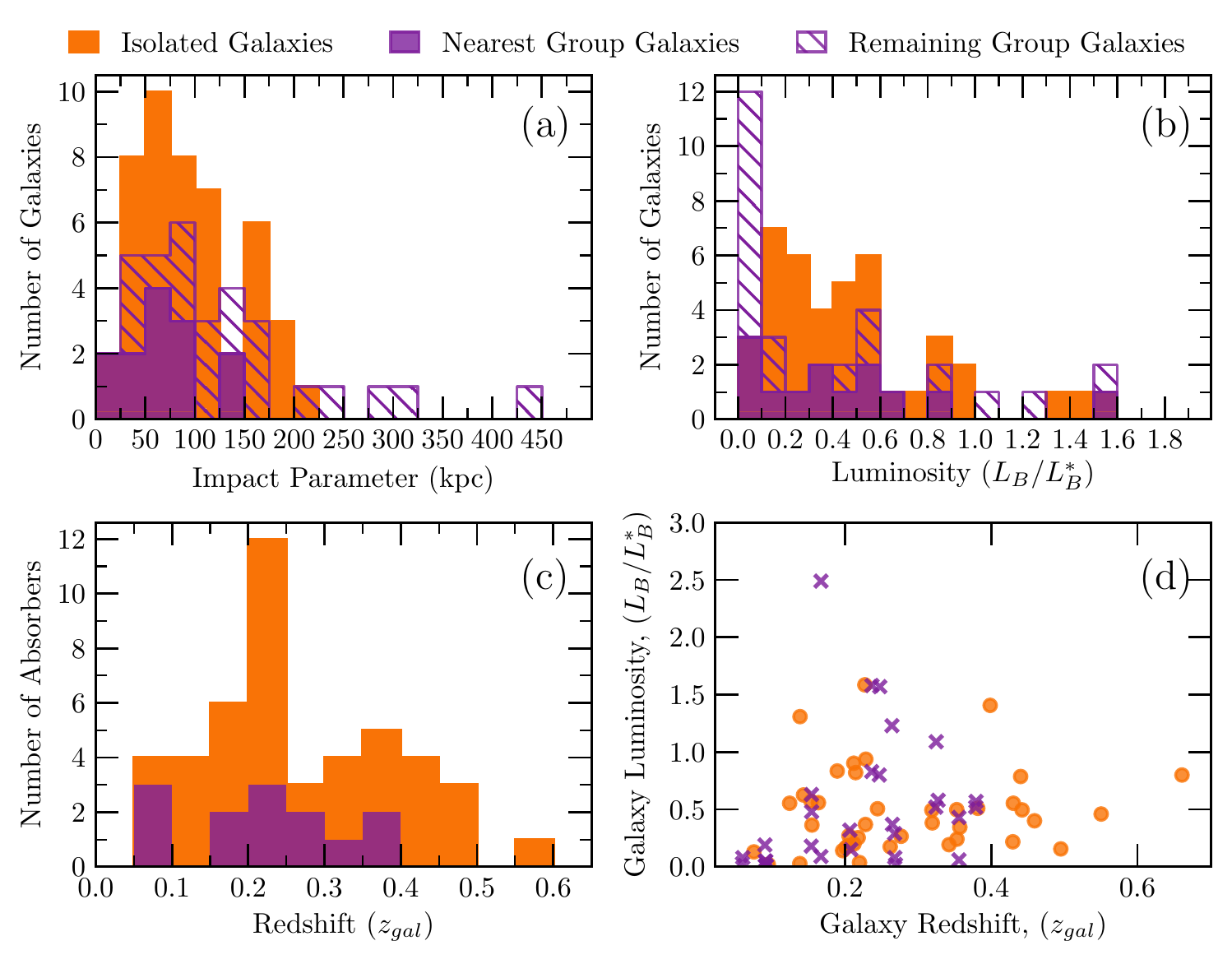}
	\caption{The distribution of isolated galaxies (orange) compared to the distribution of the nearest group galaxy member (solid purple) and all group galaxies members (hatched purple) for impact parameter (a), luminosity (b) and redshift (c). Anderson-Darling tests show that there is no significant difference between the impact parameter ($0.4\sigma$) and luminosity ($1.8\sigma$) distributions. The galaxy luminosity as a function of redshift for isolated galaxies (orange circles) and all group members (purple crosses) are shown in (d). Our luminosity sensitivity is comparable between the group and isolated environment samples until below $z=0.4$, above which we currently do not have group environment data. Although the lack of group environments above $z=0.4$ could be due to lower luminosity sensitivity at higher redshifts, our results are not dependent on selecting galaxies at all redshifts or limiting to $z<0.4$ galaxies. }
	\label{fig:samplecomp}
\end{figure*}

The isolated and group environment samples both probe a similar range of impact parameters and luminosities as shown in Figure \ref{fig:samplecomp}(a) and (b). The isolated galaxies are orange, the nearest group galaxy members are solid purple and the  remaining group galaxy members are hatched purple. We test the null hypothesis that the group galaxies are drawn from the same population as the isolated sample with an Anderson-Darling test and find that there is no significant difference between the impact parameter ($0.4\sigma$) and luminosity ($1.8\sigma$) distributions.  The details of this test and additional Anderson-Darling tests are shown in Table \ref{tab:ADtest}.

Furthermore, we show the redshift distribution of isolated and  group  environment  absorbers  in  Figure  \ref{fig:samplecomp}(c). Isolated galaxy-absorber pairs are shown in orange, while group environment absorbers are shown in purple. Although the redshift distribution of group environments covers a smaller range than that of the isolated environments, an Anderson-Darling test cannot rule out the null hypothesis that both are drawn from the same population ($1.2\sigma$). 

The galaxy redshift and luminosity relationship for the group and isolated  environments is then compared in Figure \ref{fig:samplecomp}(d). The group and isolated environment samples cover a similar range of luminosities, although the group environments only cover a range of redshifts up to $z=0.4.$ 

We do not have galaxy groups or pairs above z=0.4, which raises the possibility that the isolated galaxies at redshifts $z>0.4$ may be group environments due to poorer luminosity sensitivity at higher redshifts. To test if this affects our results, we construct a subsample of the isolated galaxies with $z<0.4$. Anderson-Darling tests between the $z<0.4$ isolated and group environment samples find that the luminosity, total impact parameter and redshift distributions are consistent ($1.41\sigma$, $0.18\sigma$ and $0.34\sigma$, respectively).  Throughout the paper, we find that comparisons between group environments and both the full and $z<0.4$ isolated environment samples are consistent and our results are not sample dependent. 

\begin{deluxetable*}{ccccc}
	\tablecolumns{5}
	\tablewidth{0pt}
	\setlength{\tabcolsep}{0.06in}
	\tablecaption{Anderson-Darling Test Results \label{tab:ADtest}}
	\tablehead{
		\colhead{Variable}           	&
        \colhead{Anderson-Darling Test Statistic}     &
        \colhead{$p$-value}&
		\colhead{Confidence Level}               &
		\colhead{$\sigma$}  }
	\startdata
	\cutinhead{Comparison of the Isolated Sample without COS Halos galaxies with the Group Sample (see Section 2.5)}
	Metallicity, ([Si/H])	&$	2.72	$&$	0.03	$&$	96.80	$&$	2.14	$	\\
Average Impact Parameter ($D$)	&$	0.15	$&$	0.98	$&$	2.15	$&$	0.03	$	\\
Most Luminous Galaxy Impact Parameter ($D$)	&$	0.19	$&$	0.93	$&$	6.60	$&$	0.08	$	\\
Nearest Galaxy Impact Parameter ($D$)	&$	3.10	$&$	0.04	$&$	96.05	$&$	2.06	$	\\
All Galaxies Impact Parameter ($D$)	&$	0.16	$&$	0.96	$&$	3.85	$&$	0.05	$	\\
All Galaxies Luminosity ($L_{B}/L^*_{B}$)	&$	2.56	$&$	0.06	$&$	93.60	$&$	1.85	$	\\
All Absorbers redshift ($z$)	&$	1.10	$&$	0.29	$&$	70.95	$&$	1.06	$	\\[-5pt]
	\cutinhead{Comparison of the Full Isolated Sample with Group Sample}
	Metallicity, ([Si/H])	&$	3.17	$&$	0.02	$&$	98.45	$&$	2.42	$	\\
Average Impact Parameter ($D$)	&$	0.41	$&$	0.70	$&$	30.40	$&$	0.39	$	\\
Most Luminous Galaxy Impact Parameter ($D$)	&$	0.28	$&$	0.88	$&$	11.70	$&$	0.15	$	\\
Nearest Galaxy Impact Parameter ($D$)	&$	2.16	$&$	0.09	$&$	90.70	$&$	1.68	$	\\
All Galaxies Impact Parameter ($D$)	&$	0.44	$&$	0.66	$&$	33.60	$&$	0.43	$	\\
All Galaxies Luminosity ($L_{B}/L^*_{B}$)	&$	2.42	$&$	0.07	$&$	92.65	$&$	1.79	$	\\
All Absorbers redshift ($z$)	&$	1.47	$&$	0.23	$&$	76.95	$&$	1.20	$\\	[-5pt]
	\cutinhead{Comparison of the $z<0.4$ Isolated Sample with Group Sample}
	Metallicity, ([Si/H])	&$	1.81	$&$	0.06	$&$	93.65	$&$	1.86	$	\\
Average Impact Parameter ($D$)	&$	0.22	$&$	0.90	$&$	9.55	$&$	0.12	$	\\
Most Luminous Galaxy Impact Parameter ($D$)	&$	0.17	$&$	0.96	$&$	4.50	$&$	0.06	$	\\
Nearest Galaxy Impact Parameter ($D$)	&$	2.42	$&$	0.08	$&$	92.00	$&$	1.75	$	\\
All Galaxies Impact Parameter ($D$)	&$	0.25	$&$	0.86	$&$	14.05	$&$	0.18	$	\\
All Galaxies Luminosity ($L_{B}/L^*_{B}$)	&$	1.75	$&$	0.16	$&$	84.00	$&$	1.41	$	\\
All Absorbers redshift ($z$)	&$	0.37	$&$	0.73	$&$	26.70	$&$	0.34	$	\\[-10pt]
	\enddata
	
\end{deluxetable*}

\section{Analysis}
The metallicities of each group environment have been inferred using the same method describe in \citet{pointon19}. We summarize the analysis in the following section.

\subsection{Spectral Analysis}\label{sec:spec_an}
Each transition was modeled using the VPFIT software \citep{carswell14} to measure the total column density. For COS spectra, we calculated the non-Gaussian line spread function (LSF) for each absorption profile using the details in \citet{kriss11} and the corresponding lifetime position. The \textit{FUSE} data were assumed to have a Gaussian LSF and a velocity resolution of $20$~{\kms} (FWHM). For the optical data from HIRES and UVES, we assumed a Gaussian LSF and a velocity resolution of $6.6$~{\kms}.

We searched for and identified up to 40 different ionic transitions within $\pm400$~{\kms} of the median redshift of the galaxy group members. We required each absorption system to have measurable {\HI} absorption features, while additional metal lines had to have reasonably consistent kinematic structure. That is, it is expected that {\MgII} absorption should have similar velocity structures to {\SiII} absorption profiles, though not necessarily to higher ionization lines which could arise in different phases. Where velocity profiles were unsaturated and uncontaminated by other absorption features, we fit one or more Voigt components to the absorption profile. To ensure that we did not over-fit the spectra, we attempted to minimize the reduced chi-squared value. However, we also required that each component still had to maintain a reasonable Doppler parameter, because extremely broad components ($b > 100$~{\kms} for {\HI} and $b>50$~{\kms} for metals) are not physical. In some cases, this resulted in a model which was physically motivated, rather than determined by the chi-squared value. 

In some absorption profiles, blends due to either contaminating gas at other redshifts or from overlapping ions were identified. In some cases, the blends were easily recognizable due to the velocity structure of the absorption profiles of other ionic transitions. However, some blends were only apparent due to the lack of consistency between the absorption profiles of different transitions of the same ionic species. Where possible, additional Voigt profile components were added to the fit to model the blend. In some cases, it was not possible to distinguish the blended absorption from the absorption profile of interest. Instead, the total column density calculated was used as a conservative upper limit on the column density. We discuss the treatment of blends for individual systems in Figure Set 1 where we present the fits.

Many of the {\HI} absorption profiles were saturated, making it difficult to accurately determine the {\HI} column density. If some lines of the {\HI} Lyman series were unsaturated or damping wings were present in the absorption profile, it was possible to obtain an accurate column density measurement. However, in the absence of unsaturated {\HI} Lyman series transitions, there exists a degeneracy between the {\HI} column density and Doppler parameter. That is, for a particular saturated {\HI} column density, the Doppler parameter may vary. Therefore, increasing the number of fitted components for a saturated absorption profile will increase the {\HI} column density. Although it is expected that the CGM is kinematically complex, resulting in many velocity components for {\HI}, the structure cannot be determined in a saturated absorption profile. Therefore, we assume that a basic one or two component fit represents the lower limit on the {\HI} column density. Due to the lack of damping wings in the absorption profile, the upper limit on the {\HI} column density is then ${\NHI}< 19.0$~{\cms}. Absorbers with column densities above this limit have damping wings are are classified as sub-DLAs or DLAs\footnote{We follow the definition in \citet{lehner18}, \citet{wotta18} and \citet{pointon19} for the classification of {\HI} absorbers. The {\HI} column density ranges for pLLSs are $16.2 < {\NHI} < 17.2$, LLS have $17.2 \leq {\NHI} < 19.0$, sub--DLAs have $19.0 \leq {\NHI} < 20.3$ and DLAs have ${\NHI} \geq 20.3$.}. 

\begin{deluxetable}{crc}
	\tablecolumns{3}
	\tablewidth{0.95\linewidth}
	\setlength{\tabcolsep}{0.06in}
	\tablecaption{J$0228$, $z_{abs} = 0.2073$ Measured Column Densities \label{tab:Q0122_0.2119}}
	\tablehead{
		\colhead{Ion}          &
         \colhead{$\log N$~({\cms})}    &
		\colhead{$\log N$ Error ({\cms})}}
	\startdata
	{\HI}   & $15.26$   &$0.02$\\
{\CII}  & $<12.79$  &$\cdots$\\
{\CIII} & $13.89$  &$0.3$\\
{\NII}  & $<12.97$  &$\cdots$\\
{\NIII} & $13.89$  &$0.19$\\
{\NV}   & $13.53$  &$0.20$\\
{\OI}   & $<13.30$  &$\cdots$\\
{\SiII} & $<11.78$  &$\cdots$\\
{\SiIII}& $<12.99$  &$\cdots$\\
{\SiIV} & $<12.43$  &$\cdots$\\[-5pt]
	\enddata
	\tablecomments{Table 3 is published in its entirety in the electronic 
edition of the {\it Astrophysical Journal}.  A portion is shown here 
for guidance regarding its form and content. The full version contains
all 13 sources.}
\end{deluxetable}

Where metal transitions were saturated, we used the fit to the profile as a lower limit on the column density. If no metal absorption was detectable, we calculated $3\sigma$ upper limits on the column density using a single cloud with an assumed Doppler parameter of $b\sim8$~{\kms}, derived from the average {\SiII} Doppler parameter. \citet{pointon19} found no significant impact on the metallicity if a larger Doppler parameter ($b=30$~{\kms}) was used.

\begin{figure*}
    \centering
    \includegraphics[width=\linewidth]{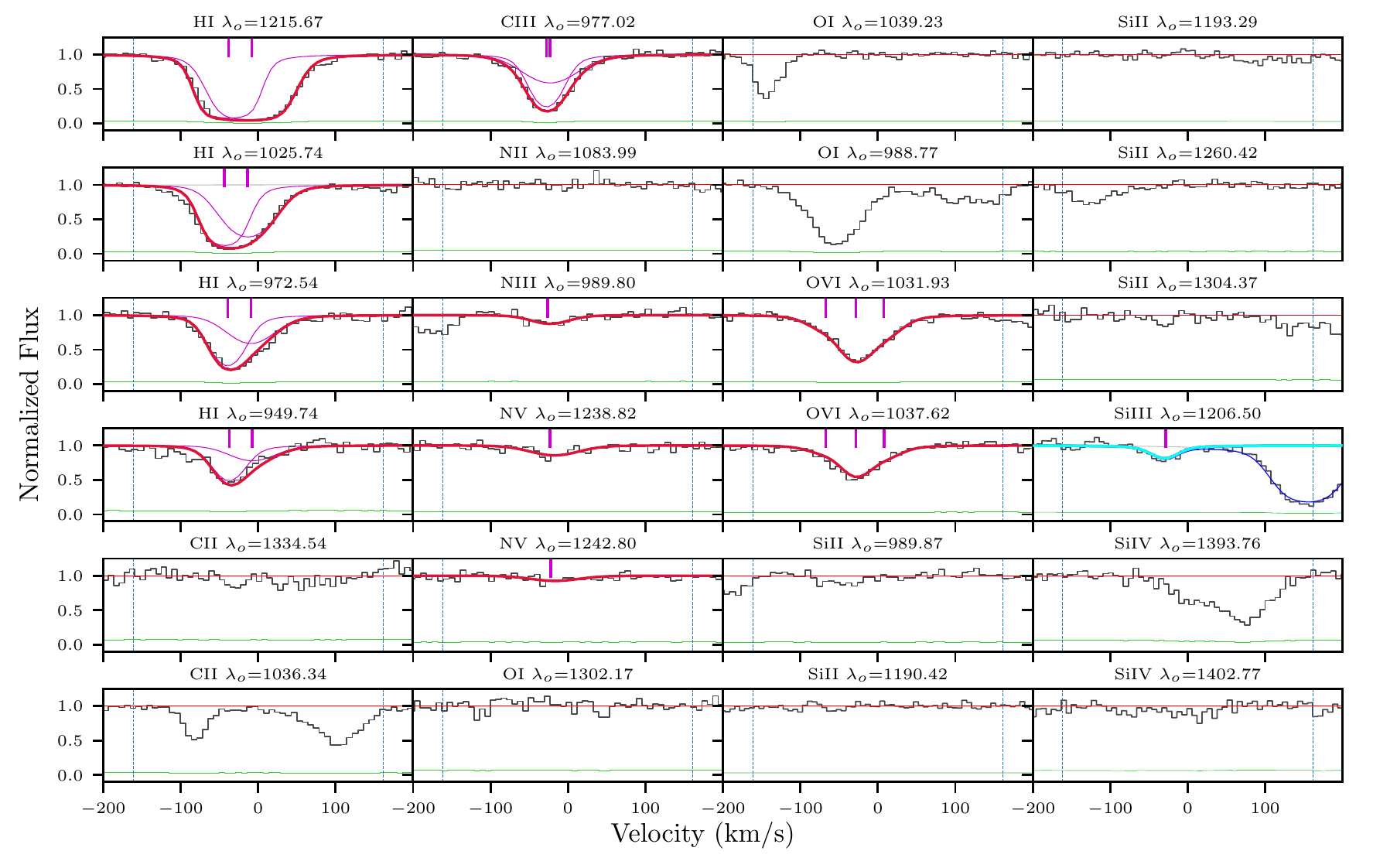}
    \caption{The fits for J$0228$, $z_{abs} = 0.2073$. The data for each ion (labelled above each panel) are shown in black, while the error spectrum is green. The fits to the absorption profiles are shown in red while the components are shown in pink. The centers of each Voigt profile used to fit the absorption profile are marked with a pink tick. The zero-point of the velocity is defined by the average redshift of the group galaxy members, while the redshifts of the galaxies in velocity space are the blue vertical lines. For ions where we calculate limits, we show the continuum level as a thin red line. It was unclear whether the {\SiIII} transition was real or a part of the complex of lines on the positive side of the spectra. Therefore, we have assumed that the column density from the {\SiIII} fit is an upper limit and it is shown in cyan. The total {\OVI} fit is shown from \citet{pointon17} for completeness, but is not used in the models. Plots for the rest of the sample are shown in Figure Set 1.}
    \label{fig:Q0122_0.2119}
\end{figure*}

We show the results of the fitting analysis in Figure \ref{fig:Q0122_0.2119} for absorption associated with the galaxy group J$0228$, $z_{abs} = 0.2073$. The black line represents the data, the green line is the error spectrum and the red line shows the fit to the absorption profiles for the ionic transition labeled above the plot. The pink lines indicate the individual components used in the fit while the pink ticks indicate the central position of each component. The redshifts of the galaxy group members are marked by vertical blue dashed lines. The velocity zero-point is defined as the average redshift of the galaxy group members. The column density measurements and limits are in Table \ref{tab:Q0122_0.2119} for J$0228$, $z_{abs} = 0.2073$. The plots of the fits and the column density data for the remaining 12 galaxy groups are shown in Figure Set 1 and the machine readable table.

\figsetstart
\figsetnum{1}
\figsettitle{The fits to each system}

\figsetgrpstart
\figsetgrpnum{1.2}
\figsetgrptitle{J$0125$, $z_{abs} = 0.3790$}
\figsetplot{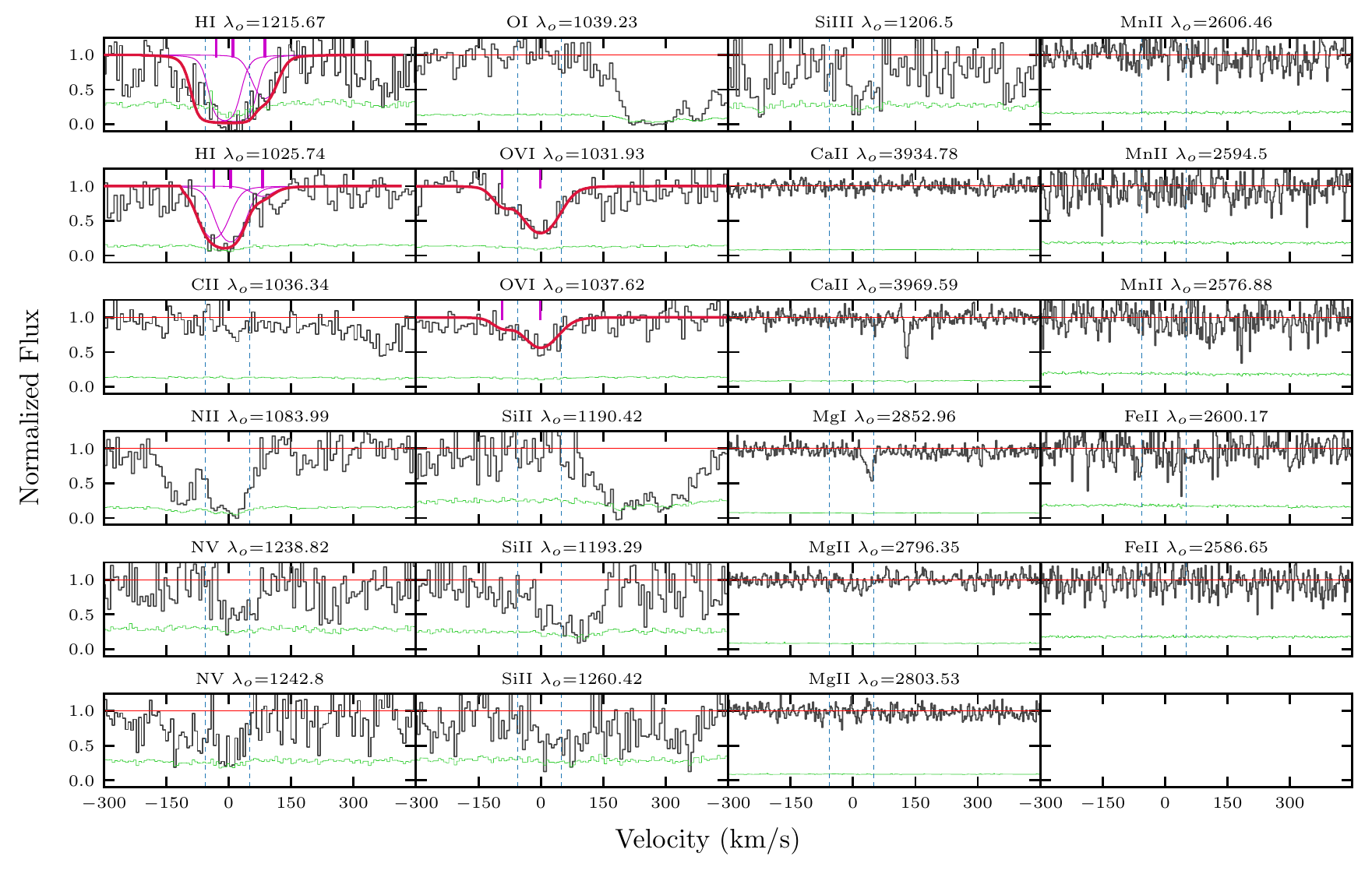}
\figsetgrpnote{The fits for J$0125$, $z_{abs} = 0.3790$. Line colors and styles are plotted as described in Figure \ref{fig:Q0122_0.2119}. We calculate upper limits for {\CII}, {\NII}, {\NV}, {\OI}, {\SiII}, {\SiIII}, {\CaII}, {\MgI}, {\MgII}, {\MnII} and {\FeII}. Note that given the possible blending on the right of the {\NV} $1238$~{\AA} line and on the left of the {\NV} $1242$~{\AA} transition it was difficult to determine a fit. Therefore we chose to calculate an upper limit.}
\figsetgrpend

\figsetgrpstart
\figsetgrpnum{1.3}
\figsetgrptitle{J$0228$, $z_{abs} = 0.2677$}
\figsetplot{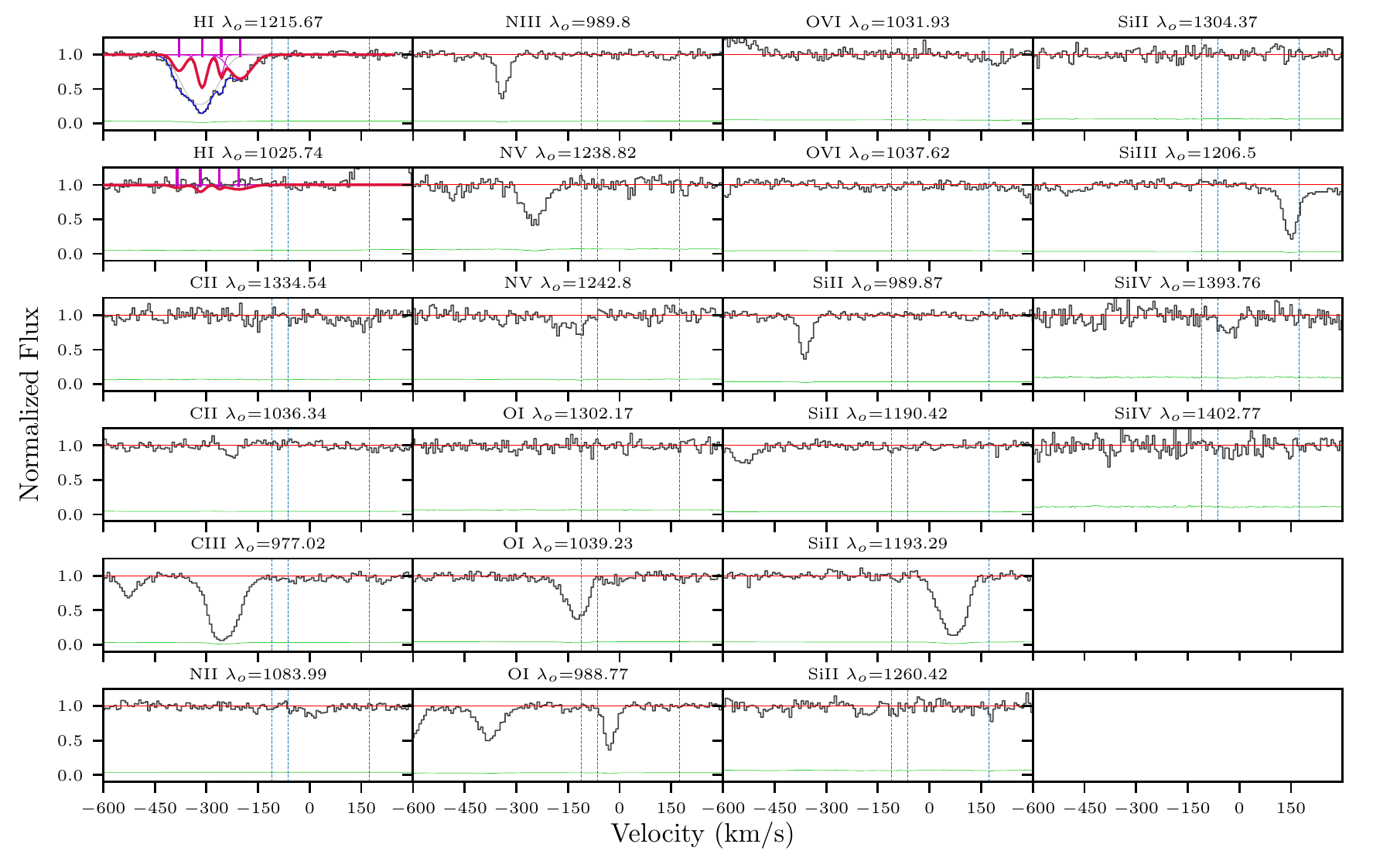}
\figsetgrpnote{The fits for J$0228$, $z_{abs} = 0.2677$. Line colors and styles are plotted as described in Figure \ref{fig:Q0122_0.2119}. Blend components are shown in grey while the total fit, consisting of both the ion and blends is shown in blue. Here the {\HI} $1215$~{\AA} line is blended with an unknown absorption system. However, the column density is well constrained from the {\HI} $1025$~{\AA} line.}
\figsetgrpend

\figsetgrpstart
\figsetgrpnum{1.4}
\figsetgrptitle{J$0351$, $z_{abs} = 0.3251$}
\figsetplot{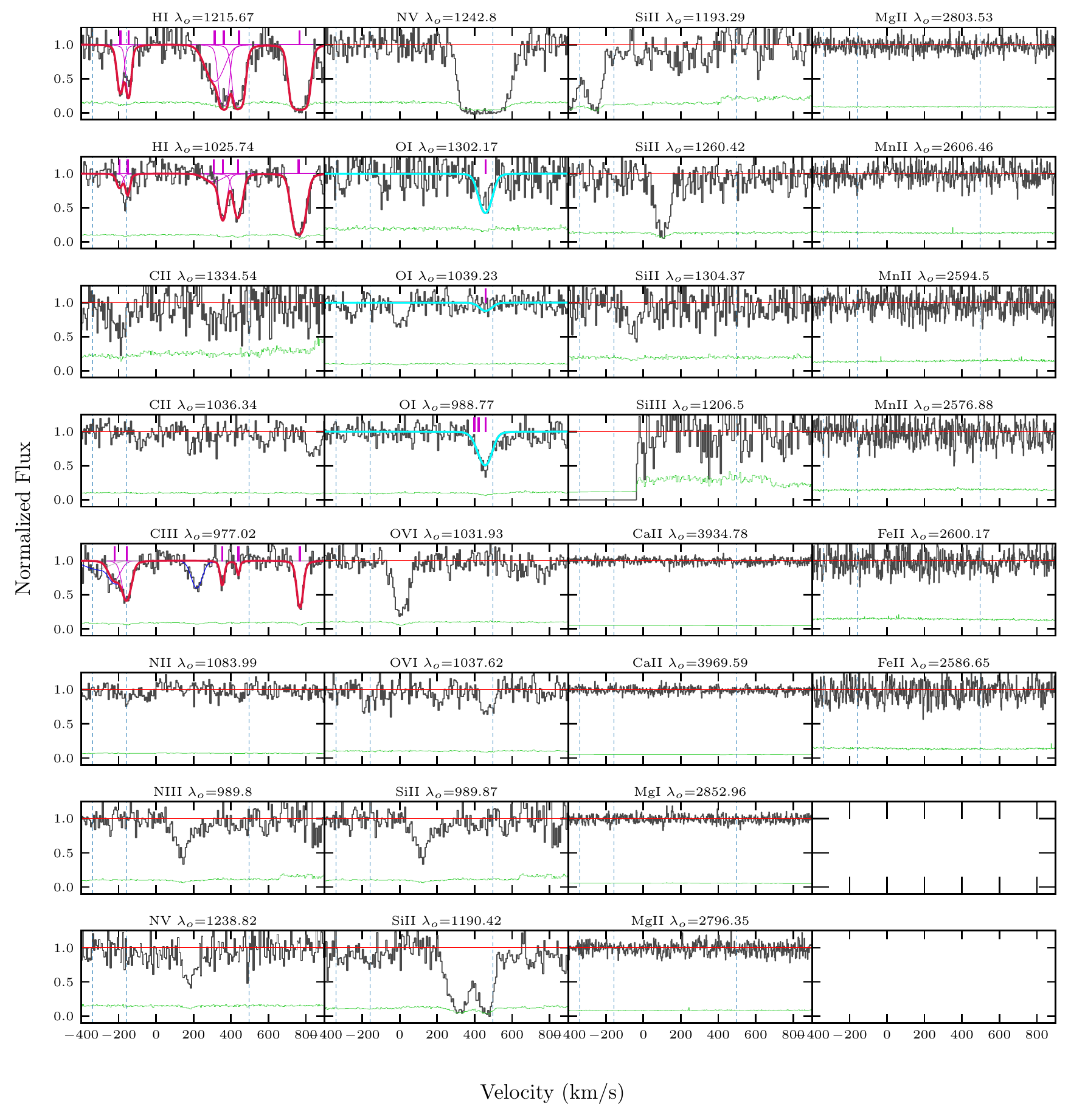}
\figsetgrpnote{The fits for J$0351$, $z_{abs} = 0.3251$. Line colors and styles are plotted as described in Figure \ref{fig:Q0122_0.2119}. Where additional components were added to the fit to de-blend the absorption profile, the total fit is shown in blue with each additional component represented by a grey line. The blend present in the {\CIII} $977$~{\AA} transition has little overlap with the modelled absorption. The {\HI} $972$~{\AA} line is blended. However, the column density is well constrained from the other {\HI} lines. The {\OI} column density obtained from the fit is treated as an upper limit.}
\figsetgrpend

\figsetgrpstart
\figsetgrpnum{1.5}
\figsetgrptitle{J$0407$, $z_{abs} = 0.0914$}
\figsetplot{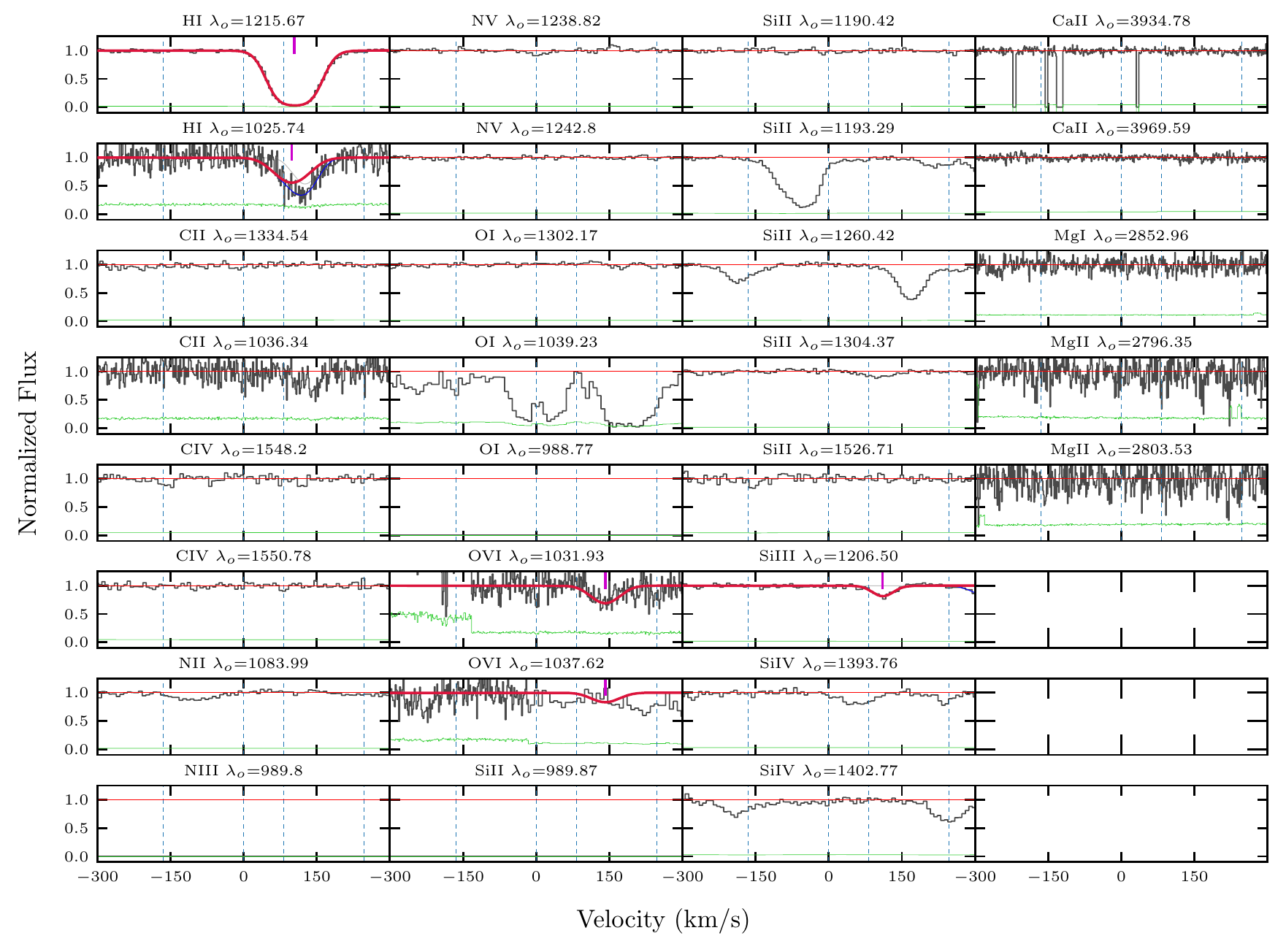}
\figsetgrpnote{The fits for J$0407$, $z_{abs} = 0.0914$. Line colors and styles are plotted as described in Figure \ref{fig:Q0122_0.2119}. Where additional components were added to the fit to de-blend the absorption profile, the total fit is shown in blue with each additional component represented by a grey line. The {\HI} $1025$~{\AA} transition has an unidentified blend. However, the {\HI} column density is constrained by the {\HI} $1215$~{\AA} transition. We have coverage for the {\HI} $1025$~{\AA}, {\CII} $1036$~{\AA} and {\OVI} lines from FUSE data.}
\figsetgrpend

\figsetgrpstart
\figsetgrpnum{1.5}
\figsetgrptitle{J$0853$, $z_{abs} = 0.0909$}
\figsetplot{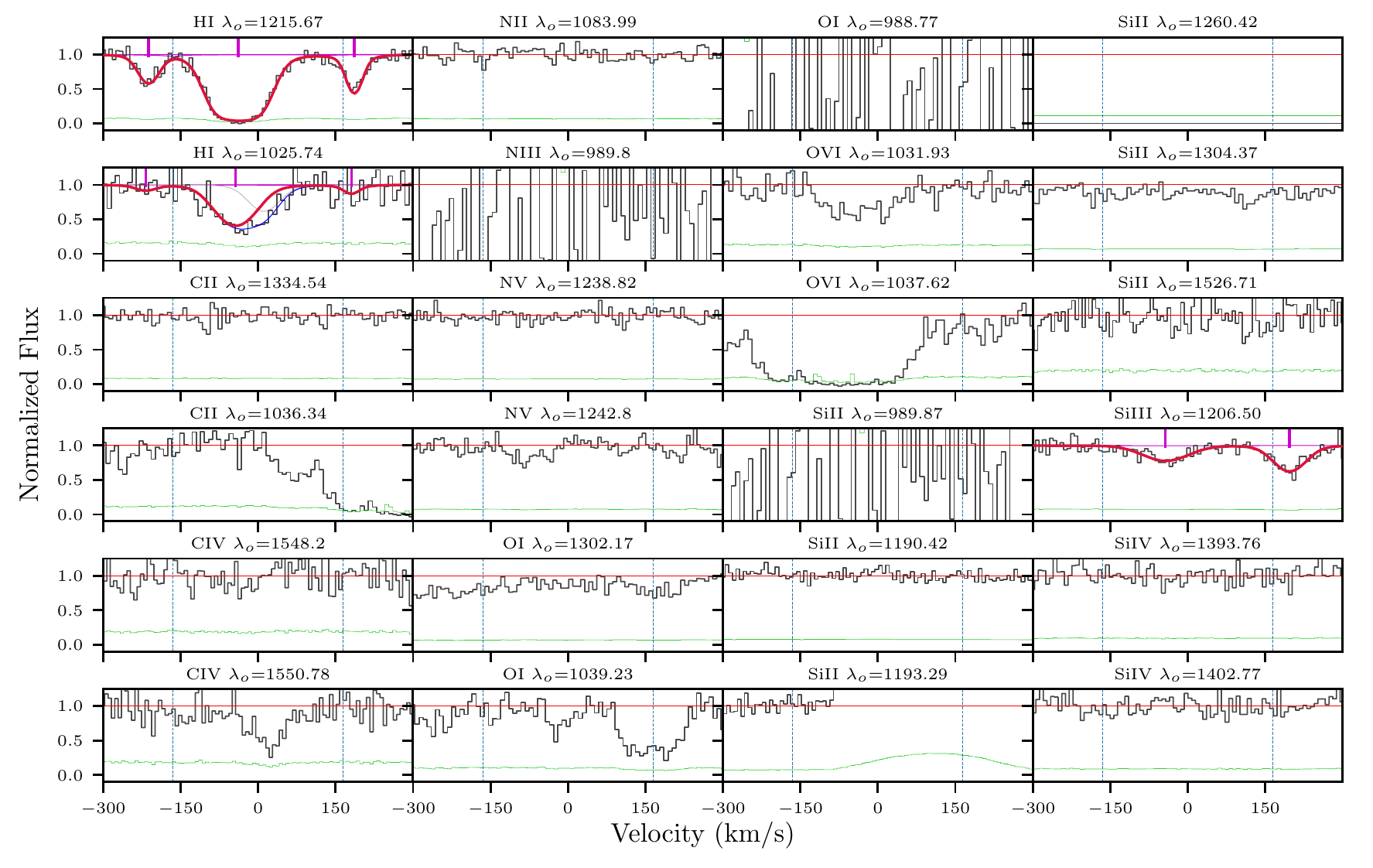}
\figsetgrpnote{The fits for J$0853$, $z_{abs} = 0.0909$. Line colors and styles are plotted as described in Figure \ref{fig:Q0122_0.2119}. Where additional components were added to the fit to de-blend the absorption profile, the total fit is shown in blue with each additional component represented by a grey line. The {\HI} $1025$~{\AA} transition has an unidentified blend. However, the {\HI} column density is constrained by the {\HI} $1215$~{\AA} transition.}
\figsetgrpend

\figsetgrpstart
\figsetgrpnum{1.6}
\figsetgrptitle{J$0910$, $z_{abs} = 0.2644$}
\figsetplot{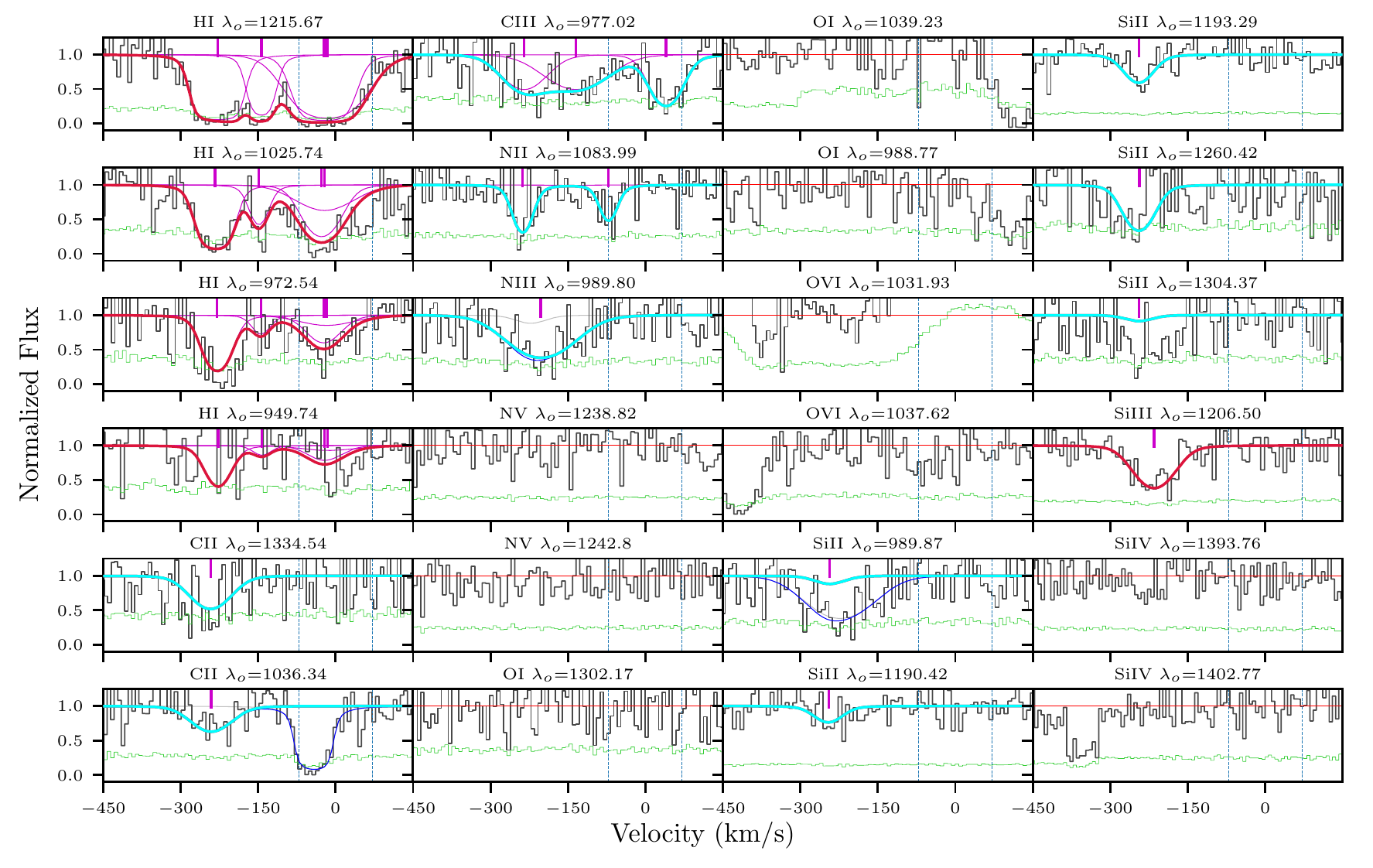}
\figsetgrpnote{The fits for J$0910$, $z_{abs} = 0.2644$. Line colors and styles are plotted as described in Figure \ref{fig:Q0122_0.2119}. Cyan line indicate ions for which we assume an upper limit on the column density in the MCMC analysis. We make this assumption due to the noise in the spectra which prevented us from obtaining reasonable column densities and Doppler parameters. Additionally, the velocity structure across many of the ions is inconsistent. Therefore, we use the column densities calculated in the fits as upper limits.}
\figsetgrpend

\figsetgrpstart
\figsetgrpnum{1.7}
\figsetgrptitle{J$0925$, $z_{abs} = 0.2471$}
\figsetplot{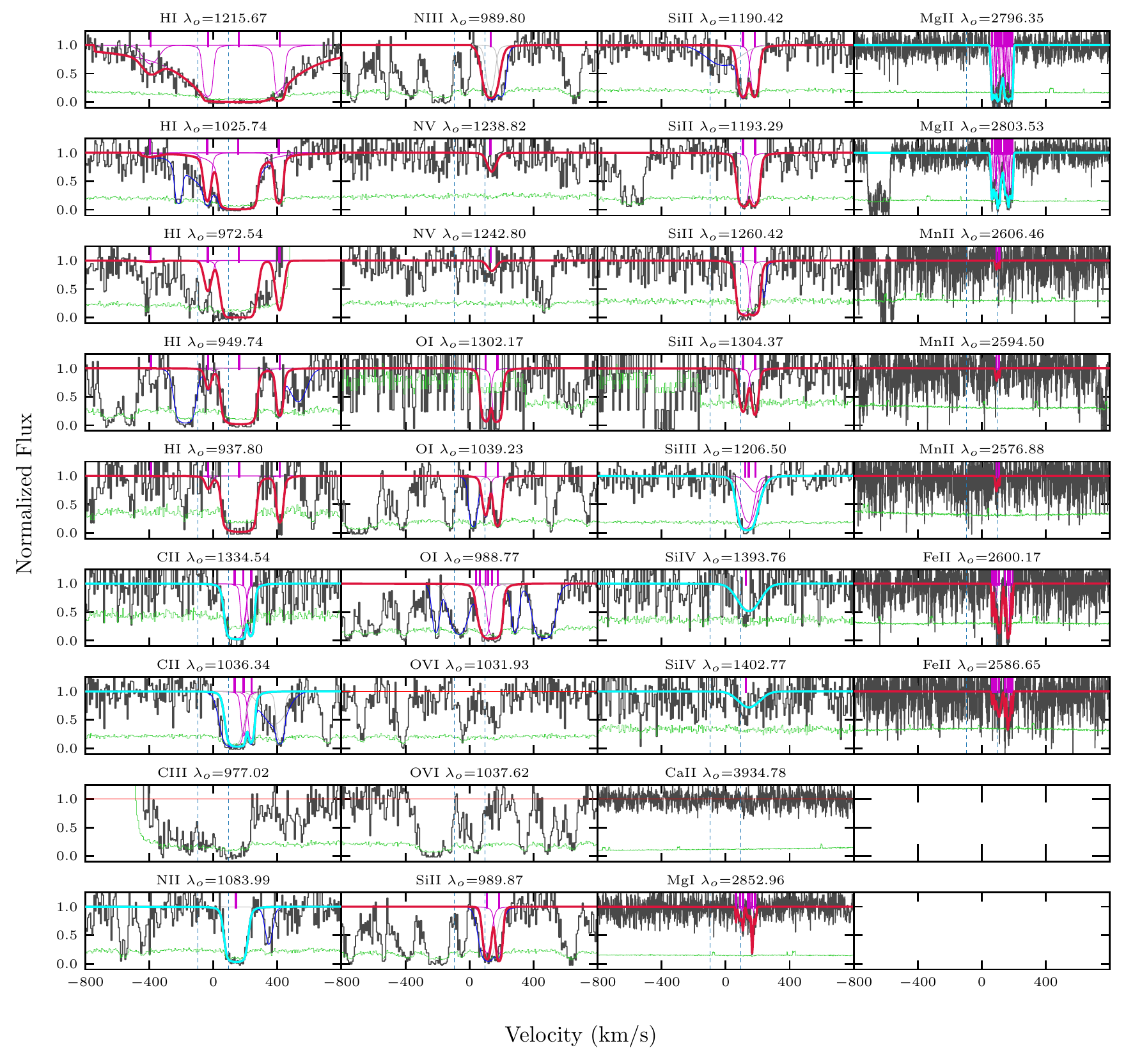}
\figsetgrpnote{The fits for J$0925$, $z_{abs} = 0.2471$. Line colors and styles are plotted as described in Figure \ref{fig:Q0122_0.2119}. Where additional components were added to the fit to de-blend the absorption profile, the total fit is shown in blue with each additional component represented by a grey line. There are unidentified blends in the {\HI} $1025$~{\AA}, {\CII} $1036$~{\AA} , {\OI} $1039$~{\AA}, {\OI} $988$~{\AA}, {\SiII} $989$~{\AA}, {\SiII} $1190$~{\AA} and {\SiII} $1260$~{\AA} transitions. However, for these ions, the column densities can be constrained using other unblended transitions of the same ionization state. The blend in the {\NII} transition has minimal overlap with the absorption. {\NIII} $989$~{\AA} and {\SiII} $989$~{\AA} are blended together. However the {\SiII} column density is constrained by other {\SiII} lines. We assume that the remaining absorption of from {\NIII}. Due to saturation we assumed the measured column densities for {\CII}, {\NII}, {\SiIII} and {\MgII} are lower limits. Due to noise in the spectrum, we are only able to determine that the {\SiIV} column density is an upper limit. Where the measured column density was used as limit, the fit is shown in cyan. We plot the modelled {\HI} $972$~{\AA} absorption, although we do not use the data to constrain the fit due to the strong emission feature.}
\figsetgrpend

\figsetgrpstart
\figsetgrpnum{1.8}
\figsetgrptitle{J$0928$, $z_{abs} = 0.1540$}
\figsetplot{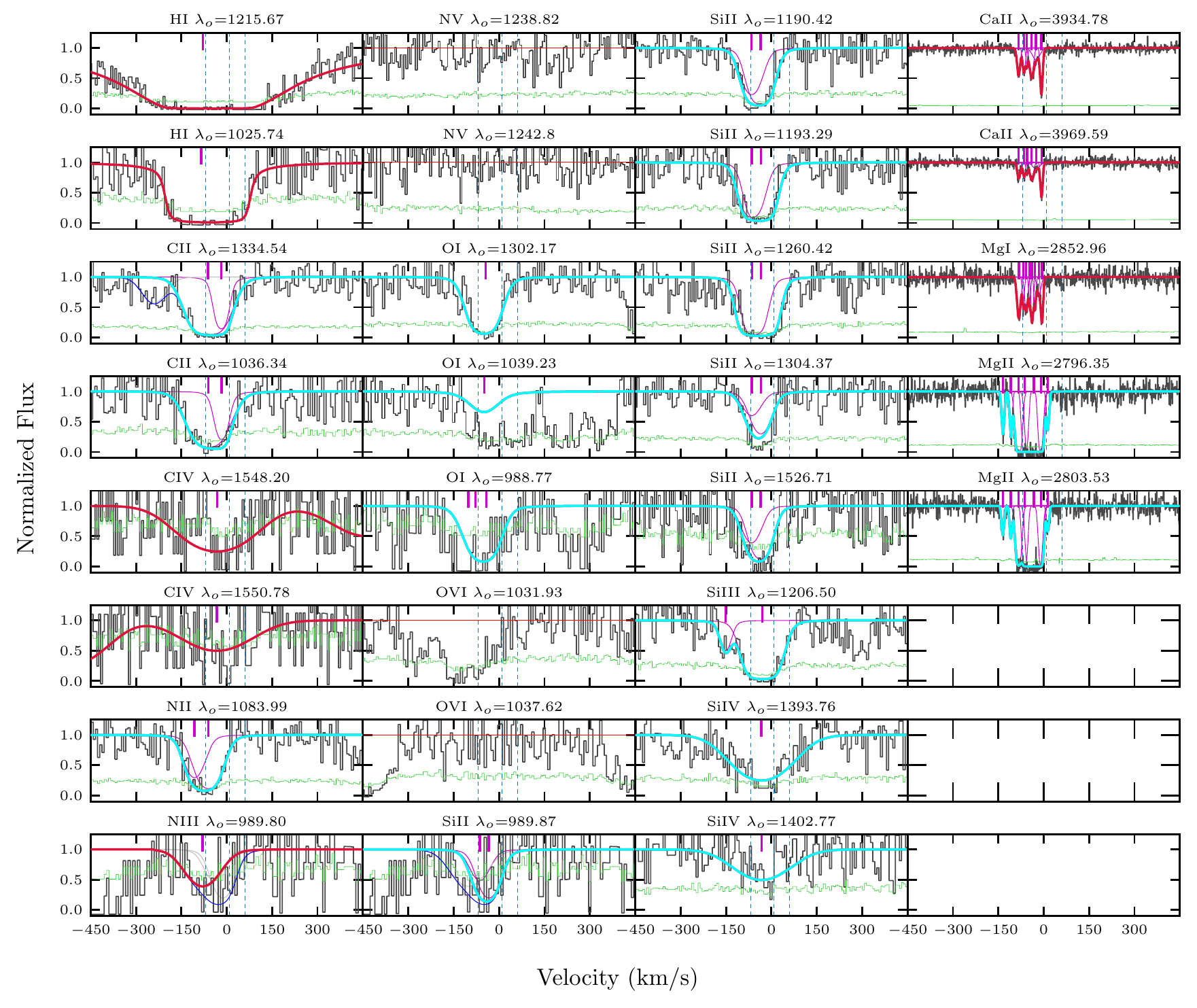}
\figsetgrpnote{The fits for J$0928$, $z_{abs} = 0.1540$. Line colors and styles are plotted as described in Figure \ref{fig:Q0122_0.2119}. Where additional components were added to the fit to de-blend the absorption profile, the total fit is shown in blue with each additional component represented by a grey line. Although the {\NIII} $989$~{\AA} and {\SiII} $989$~{\AA} are blended together, the {\SiII} column density is constrained by other {\SiII} lines. We then assume that the remaining absorption is from {\NIII}. Due to saturation, we have assumed that the column densities measured for {\CII}, {\NII}, {\OI}, {\SiII}, {\SiIII} and {\MgII} are upper limits. Furthermore, due to noise in the spectrum, we were unable to find a good fit to the {\SiIV} transition and take the measured column density as an upper limit. Where the measured column density was used as an upper limit, the fit is shown in cyan.}
\figsetgrpend

\figsetgrpstart
\figsetgrpnum{1.9}
\figsetgrptitle{J$1009$, $z_{abs} = 0.3556$}
\figsetplot{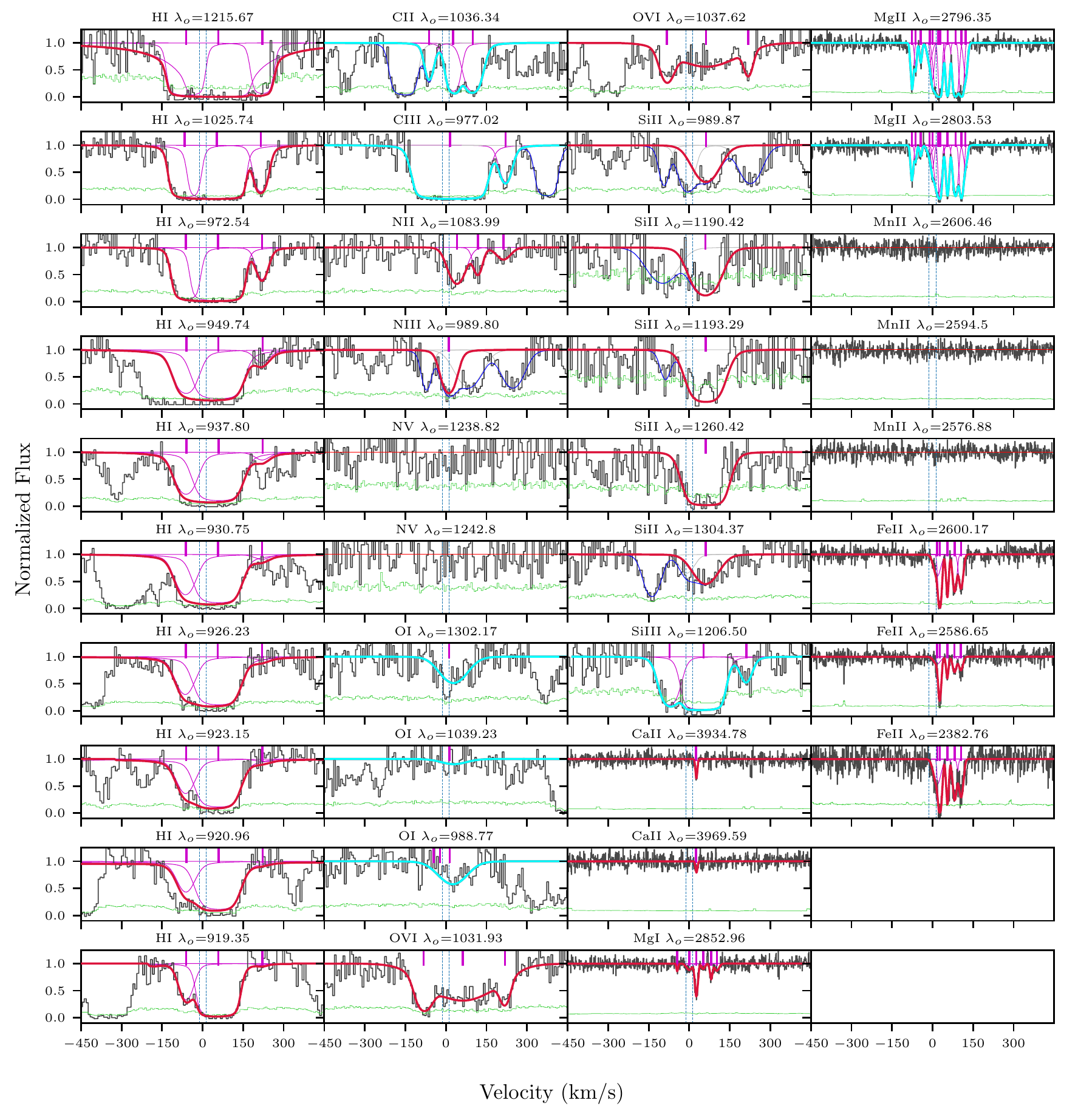}
\figsetgrpnote{The fits for J$1009$, $z_{abs} = 0.3556$. Line colors and styles are plotted as described in Figure \ref{fig:Q0122_0.2119}. Where additional components were added to the fit to de-blend the absorption profile, the total fit is shown in blue with each additional component represented by a grey line. In cases were the measured column density is used as a limit, we plot the fit in cyan. Although the {\NIII} $989$~{\AA} and {\SiII} $989$~{\AA} are blended together, the {\SiII} column density is constrained by other {\SiII} lines. We assume that the remaining absorption is from {\NIII}. The blends in the {\CII} $1036$~{\AA}, {\CIII} $977$~{\AA} and {\SiII} $1190$~{\AA}, $1193$~{\AA} and $1304$~{\AA} transitions either have minimal overlap with the absorption or the column density is constrained from another transition. Due to saturation, we assumed the measured column densities from {\CII}, {\CIII}, {\SiIII} and {\MgII} are lower limits. We also note that the {\OI} transition is unusually strong, hence we assumed the measured column density is an upper limit.}
\figsetgrpend

\figsetgrpstart
\figsetgrpnum{1.10}
\figsetgrptitle{J$1119$, $z_{abs} = 0.0600$}
\figsetplot{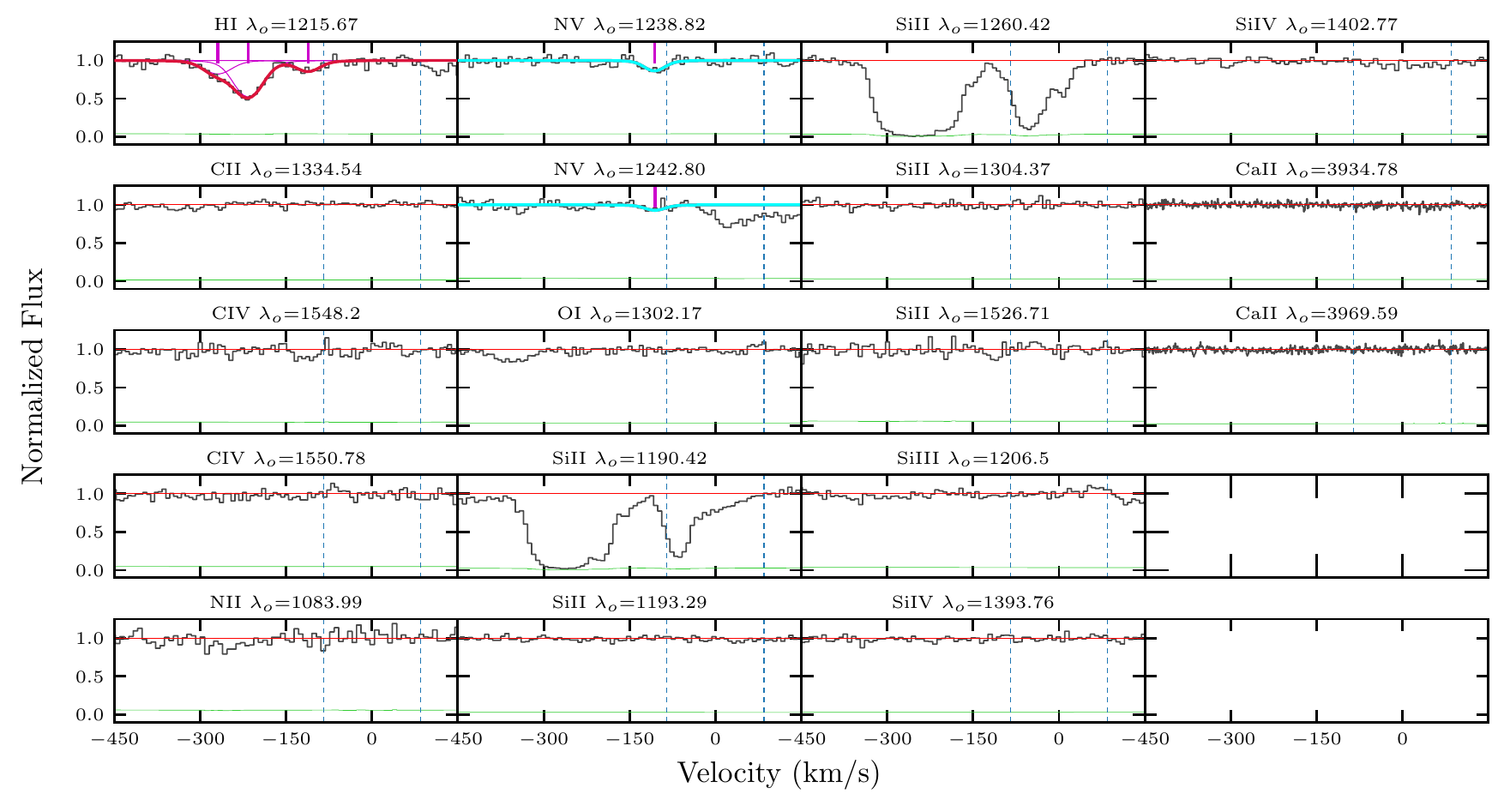}
\figsetgrpnote{The fits for J$1119$, $z_{abs} = 0.0600$. Line colors and styles are plotted as described in Figure \ref{fig:Q0122_0.2119}. We assume an upper limit for {\NV} since the detection is on the level of the noise in the spectrum.}
\figsetgrpend

\figsetgrpstart
\figsetgrpnum{1.11}
\figsetgrptitle{J$1133$, $z_{abs} = 0.2366$}
\figsetplot{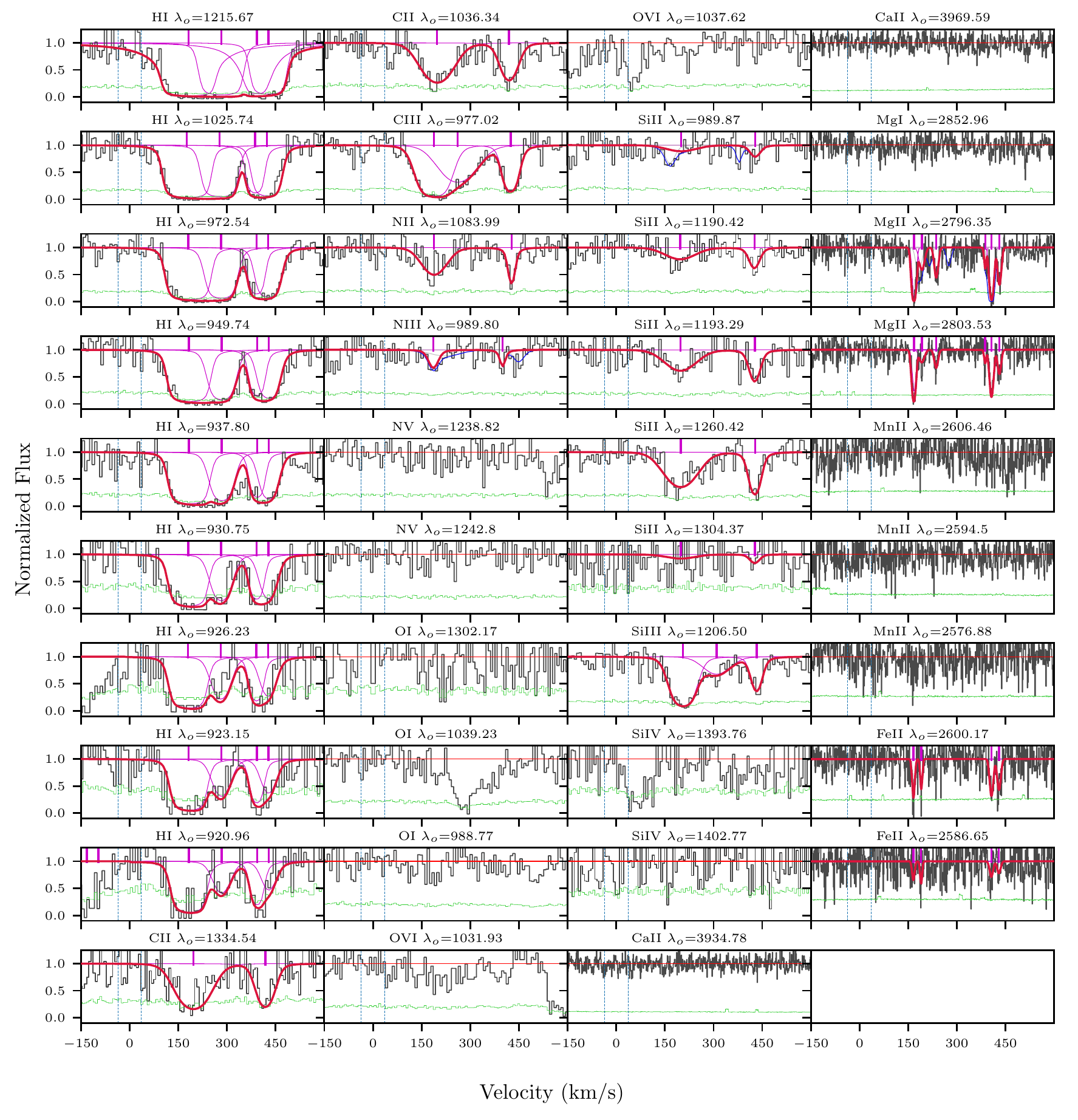}
\figsetgrpnote{The fits for J$1133$, $z_{abs} = 0.2366$. Line colors and styles are plotted as described in Figure \ref{fig:Q0122_0.2119}. Where additional components were added to the fit to de-blend the absorption profile, the total fit is shown in blue with each additional component represented by a grey line. {\NIII} $989$~{\AA} and {\SiII} $989$~{\AA} are blended together. However the {\SiII} column density is constrained by other {\SiII} lines. We assume that the remaining absorption is from {\NIII}. The {\CIII} transition is saturated, resulting an an upper limit on the column density.}
\figsetgrpend


\figsetend

The analysis method used enables the determination of column densities for {\HI} Lyman series, {\CII}, {\CIII}, {\CIV}, {\NII}, {\NIII}, {\NV}, {\OI}, {\SiII}, {\SiIII}, {\SiIV}, {\CaII}, {\MgI}, {\MgII} and {\FeII}, which are then applied in the ionization modelling to determine the metallicity of the CGM. We note that the {\OVI} column densities are presented in \citet{pointon17} and the fits are shown in this work for completeness.

\subsection{Ionization Modelling}

\begin{deluxetable*}{cccrrrr}
	\tablecolumns{7}
	\tablewidth{0pt}
	\setlength{\tabcolsep}{0.06in}
	\tablecaption{MCMC Output \label{tab:ionization_param}}
	\tablehead{
		\colhead{J-Name}           	&
        \colhead{$z_{\rm abs}$}     &
        \colhead{Meas. ${\NHI}$\tablenotemark{a}}&
		\colhead{[Si/H]\tablenotemark{b}}               &
		\colhead{${\NHI}$\tablenotemark{b}}    &
		\colhead{$\log n_{\tiny H}$\tablenotemark{b}}	&
		\colhead{$\log U$\tablenotemark{b}}\\
		\colhead{}           	&
        \colhead{}     &
        \colhead{({\cms})}&
		\colhead{}               &
		\colhead{({\cms})}    &
		\colhead{($\text{cm}^{-3}$)}	&
		\colhead{}}
	\startdata
	J0125	&	$	0.3790	$	&	$	15.48	\pm	0.02	$	&	$	<	0.06						$	&	$	15.12	_{	0.17	}^{	0.05	}	$	&	$	<	-2.003						$	&	$	<	-1.27						$	\\[2pt]
J0228	&	$	0.2073	$	&	$	15.26	\pm	0.02	$	&	$		-1.55	_{	-0.12	}^{	-0.07	}	$	&	$	15.26	_{	0.02	}^{	0.01	}	$	&	$		-4.246	_{	0.025	}^{	0.084	}	$	&	$		-1.45	_{	0.01	}^{	0.03	}	$	\\[2pt]
J0228	&	$	0.2677	$	&	$	14.21	\pm	0.01	$	&	$	<	0.65						$	&	$	14.21	_{	0.01	}^{	0.01	}	$	&	$	<	-1.101						$	&	$	<	-1.35						$	\\[2pt]
J0351	&	$	0.3251	$	&	$	15.26	\pm	0.02	$	&	$		-1.39	_{	-0.37	}^{	-0.56	}	$	&	$	15.26	_{	0.02	}^{	0.02	}	$	&	$		-3.779	_{	0.389	}^{	0.623	}	$	&	$		-1.91	_{	0.20	}^{	0.32	}	$	\\[2pt]
J0407	&	$	0.0914	$	&	$	14.36	\pm	0.01	$	&	$		-0.18	_{	-0.14	}^{	-0.06	}	$	&	$	14.36	_{	0.01	}^{	0.01	}	$	&	$		-3.384	_{	0.094	}^{	0.094	}	$	&	$		-2.43	_{	0.07	}^{	0.07	}	$	\\[2pt]
J0407	&	$	0.1670\tablenotemark{c}	$	&	$	16.45	\pm	0.05	$	&	$		-0.10	_{	-0.02	}^{	-0.02	}	$	&	$	16.45	_{	0.05	}^{	0.05	}	$	&	$		-2.800	_{	0.060	}^{	0.060	}	$	&	$		-3.20	_{	0.06	}^{	0.06	}	$	\\[2pt]
J0853	&	$	0.0909	$	&	$	14.68	\pm	0.04	$	&	$		-0.02	_{	-0.10	}^{	-0.04	}	$	&	$	14.70	_{	0.03	}^{	0.04	}	$	&	$		-3.404	_{	0.052	}^{	0.036	}	$	&	$		-2.40	_{	0.04	}^{	0.03	}	$	\\[2pt]
J0910	&	$	0.2644	$	&	$	15.47	\pm	0.07	$	&	$		-1.04	_{	-0.31	}^{	-0.65	}	$	&	$	15.45	_{	0.17	}^{	0.11	}	$	&	$		-3.429	_{	-0.195	}^{	0.971	}	$	&	$		-3.32	_{	-0.19	}^{	0.94	}	$	\\[2pt]
J0925	&	$	0.2471	$	&	$	19.58	\pm	0.02	$	&	$		-0.77	_{	-0.02	}^{	-0.03	}	$	&	$	19.60	_{	0.02	}^{	0.02	}	$	&	$		-3.135	_{	0.001	}^{	0.016	}	$	&	$		-2.48	_{	0.01	}^{	0.01	}	$	\\[2pt]
J0928	&	$	0.1540	$	&	$	19.47	\pm	0.02	$	&	$		-0.23	_{	-0.05	}^{	-0.04	}	$	&	$	19.47	_{	0.02	}^{	0.02	}	$	&	$		-3.131	_{	0.048	}^{	0.045	}	$	&	$		-2.60	_{	0.04	}^{	0.04	}	$	\\[2pt]
J1009	&	$	0.3556	$	&	$	18.96   \pm 0.07	$	&	$		-0.84	_{	-0.04	}^{	-0.03	}	$	&	$	18.54	_{	0.03	}^{	0.04	}	$	&	$		-3.081	_{	0.000	}^{	0.014	}	$	&	$		-2.40	_{	0.01	}^{	0.01	}	$	\\[2pt]
J1119	&	$	0.0597	$	&	$	13.68	\pm	0.02	$	&	$	<	0.69						$	&	$	13.68	_{	0.02	}^{	0.02	}	$	&	$	<	-2.001						$	&	$	<	-1.13						$	\\[2pt]
J1133	&	$	0.2366	$	&	$	[18.35, 19.00]	$	&	$		-1.71	_{	-0.04	}^{	-0.03	}	$	&	$	18.50	_{	0.08	}^{	0.05	}	$	&	$		-2.919	_{	0.078	}^{	0.074	}	$	&	$		-2.69	_{	0.07	}^{	0.07	}	$	\\[-5pt]
	\enddata
	\tablenotetext{a}{{\HI} column density measured from the Voigt profile modelling of the absorption profiles.}
	\tablenotetext{b}{The most likely value with the $68\%$ uncertainties from the MCMC analysis. For upper limits, we take the $95\%$ upper uncertainty. }
	\tablenotetext{c}{Results from ionization modelling taken from \citet{muzahid18}.}
\end{deluxetable*}

A single low ionization phase metallicity for each group environment is calculated by comparing a grid of predicted column densities modeled by the ionization modeling suite Cloudy to the column densities calculated in the previous section. Cloudy uses the input ionization conditions, set by the {\HI} column density, $N_{\hbox{\tiny \HI}}$, hydrogen density, $n_{\text{H}}$ and metallicity, [Si/H], to predict the column densities of the metals in the gas \citep{ferland13}. Typical grids cover a range $-5.0 < \log n_{\text{H}} < -1.0~\text{cm}^{-3}$, $13.0 < {\NHI} < 20.0~\text{cm}^{-2}$ and $-4.0 < [\text{Si}/\text{H}] < 1.5$. We assumed a uniform layer of gas, with no dust and solar abundance ratios, is irradiated by a background UV spectrum. The gas is also assumed to be single-phase, leading to the exclusion of the highly ionized {\OVI} gas from the analysis. For consistency with \citet{lehner13}, \citet{wotta16}, \citet{wotta18} and \citet{pointon19}, we adopt the ionizing background spectrum described by the Haardt and Madau 2005 model \citep[HM05;][as implemented in Cloudy]{haardt01}. The shape of the ionizing background, which can have an impact on the metallicity \citep{fechner11}, is also assumed to only evolve with redshift.

The metallicity and ionization parameter of each absorption system are then inferred by a MCMC technique described by \citet{crighton13}. The column densities in each grid point calculated by Cloudy are compared to the measured column densities. Upper and lower limits are treated as one-sided Gaussians by the likelihood function. Priors were set to the boundaries of the Cloudy ionization grids in most cases or to the upper or lower limits of the {\HI} column density, shown in the column density tables. For each group, we initialize the MCMC analysis with 100 walkers and a burn-in of 200 steps. The final distributions of the MCMC walkers, from which we infer the metallicity and ionization parameter, are then determined by another 200 steps.

\figsetstart
\figsetnum{2}
\figsettitle{The posterior distribution profiles from the MCMC analysis}
\figsetgrpstart
\figsetgrpnum{2.1}
\figsetgrptitle{J$0125$, $z_{abs} = 0.3790$}
\figsetplot{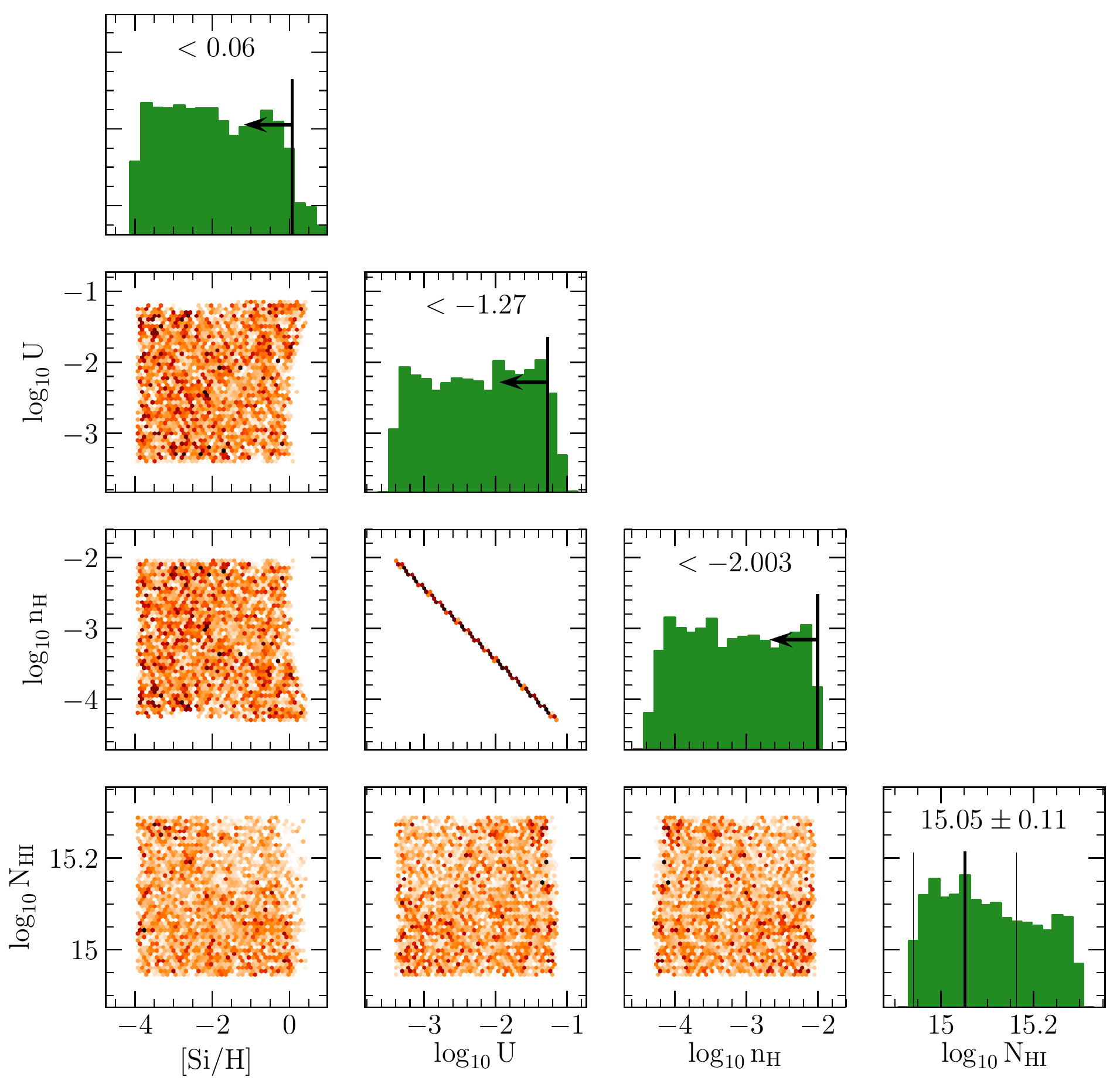}
\figsetgrpnote{The posterior distribution profiles from the MCMC analysis of the Cloudy grids for J$0125$, $z_{abs} = 0.3790$. The model parameters are plotted as described in Figure \ref{fig:Q0122_0.2119}.}
\figsetgrpend

\figsetgrpstart
\figsetgrpnum{2.1}
\figsetgrptitle{}
\figsetplot{}
\figsetgrpnote{}
\figsetgrpend

\figsetgrpstart
\figsetgrpnum{2.2}
\figsetgrptitle{J$0228$, $z_{abs} = 0.2677$}
\figsetplot{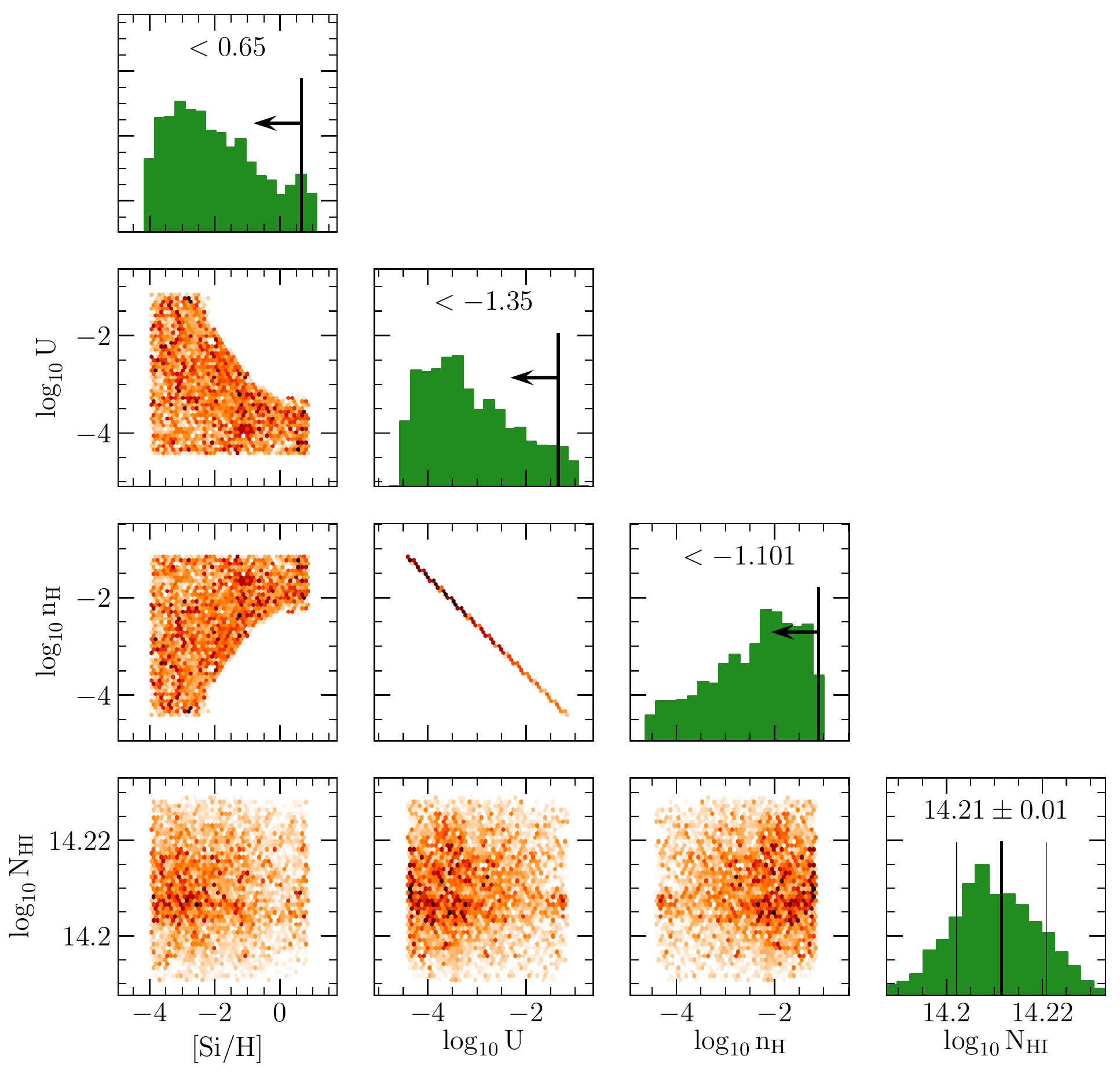}
\figsetgrpnote{The posterior distribution profiles from the MCMC analysis of the Cloudy grids for J$0228$, $z_{abs} = 0.2677$. The model parameters are plotted as described in Figure \ref{fig:Q0122_0.2119}.}
\figsetgrpend

\figsetgrpstart
\figsetgrpnum{2.3}
\figsetgrptitle{J$0351$, $z_{abs} = 0.3251$}
\figsetplot{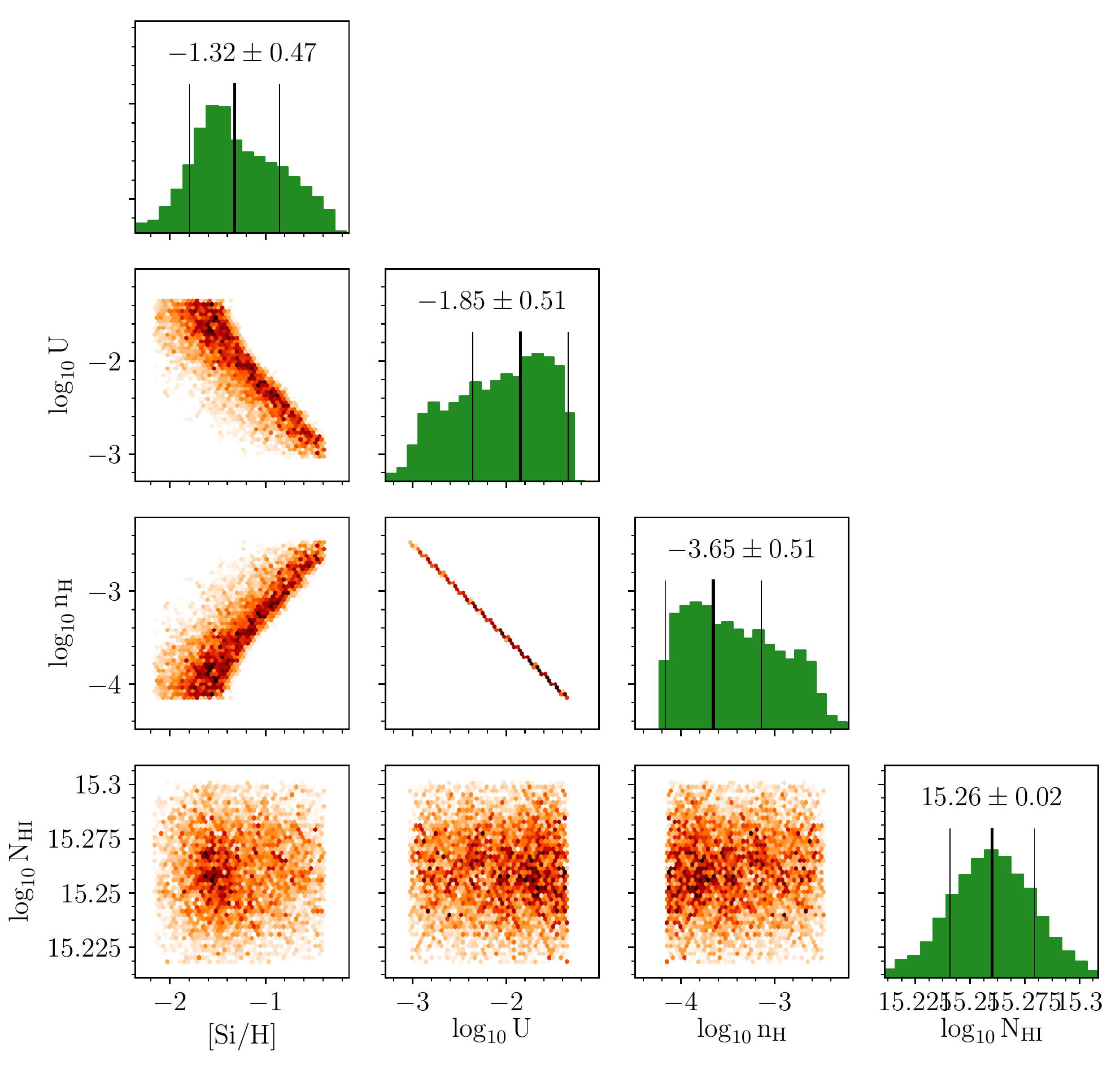}
\figsetgrpnote{The posterior distribution profiles from the MCMC analysis of the Cloudy grids for J$0351$, $z_{abs} = 0.3251$. The model parameters are plotted as described in Figure \ref{fig:Q0122_0.2119}.}
\figsetgrpend

\figsetgrpstart
\figsetgrpnum{2.4}
\figsetgrptitle{J$0407$, $z_{abs} = 0.0914$}
\figsetplot{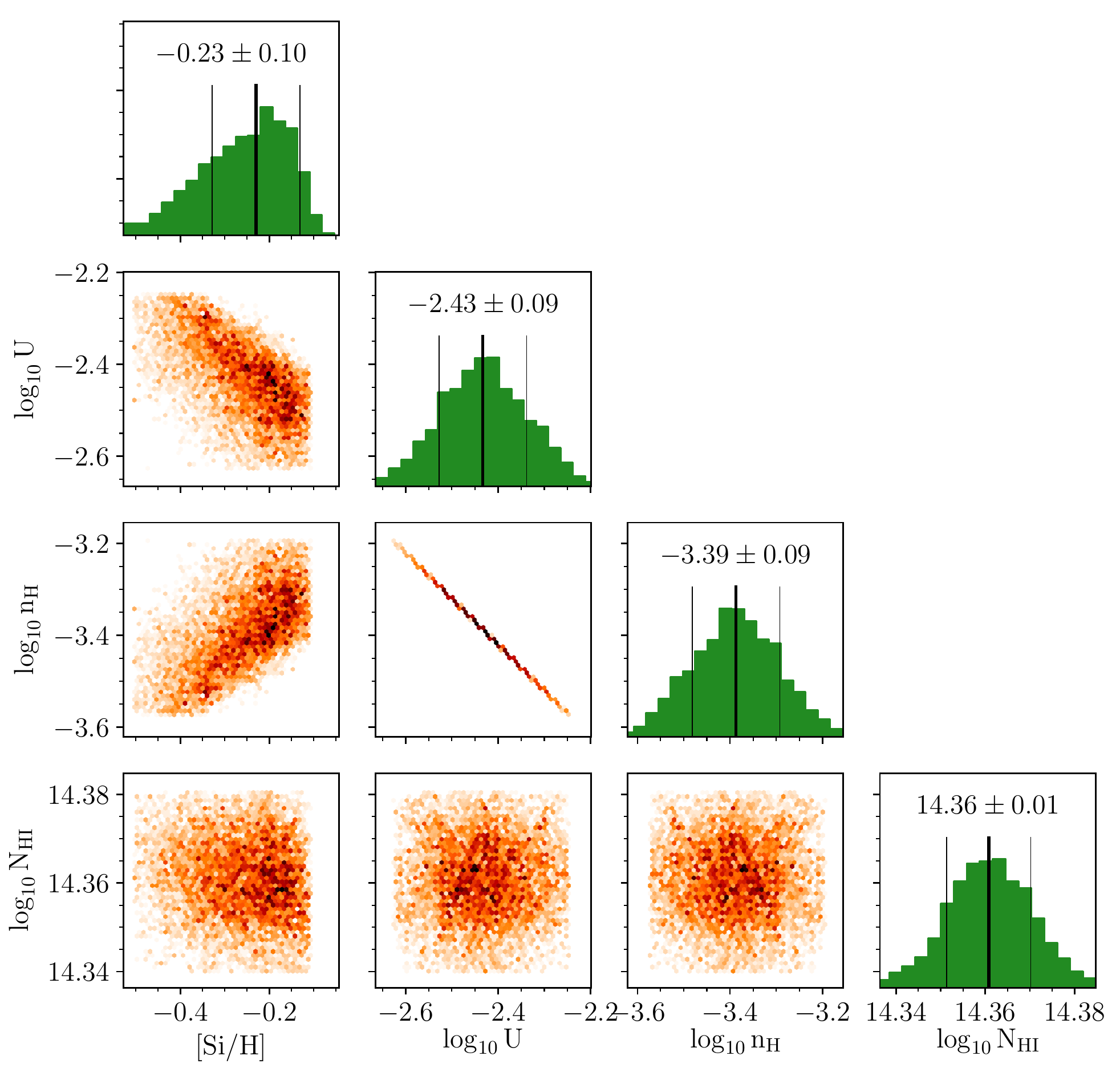}
\figsetgrpnote{The posterior distribution profiles from the MCMC analysis of the Cloudy grids for J$0407$, $z_{abs} = 0.0914$. The model parameters are plotted as described in Figure \ref{fig:Q0122_0.2119}.}
\figsetgrpend

\figsetgrpstart
\figsetgrpnum{2.5}
\figsetgrptitle{J$0853$, $z_{abs} = 0.0909$}
\figsetplot{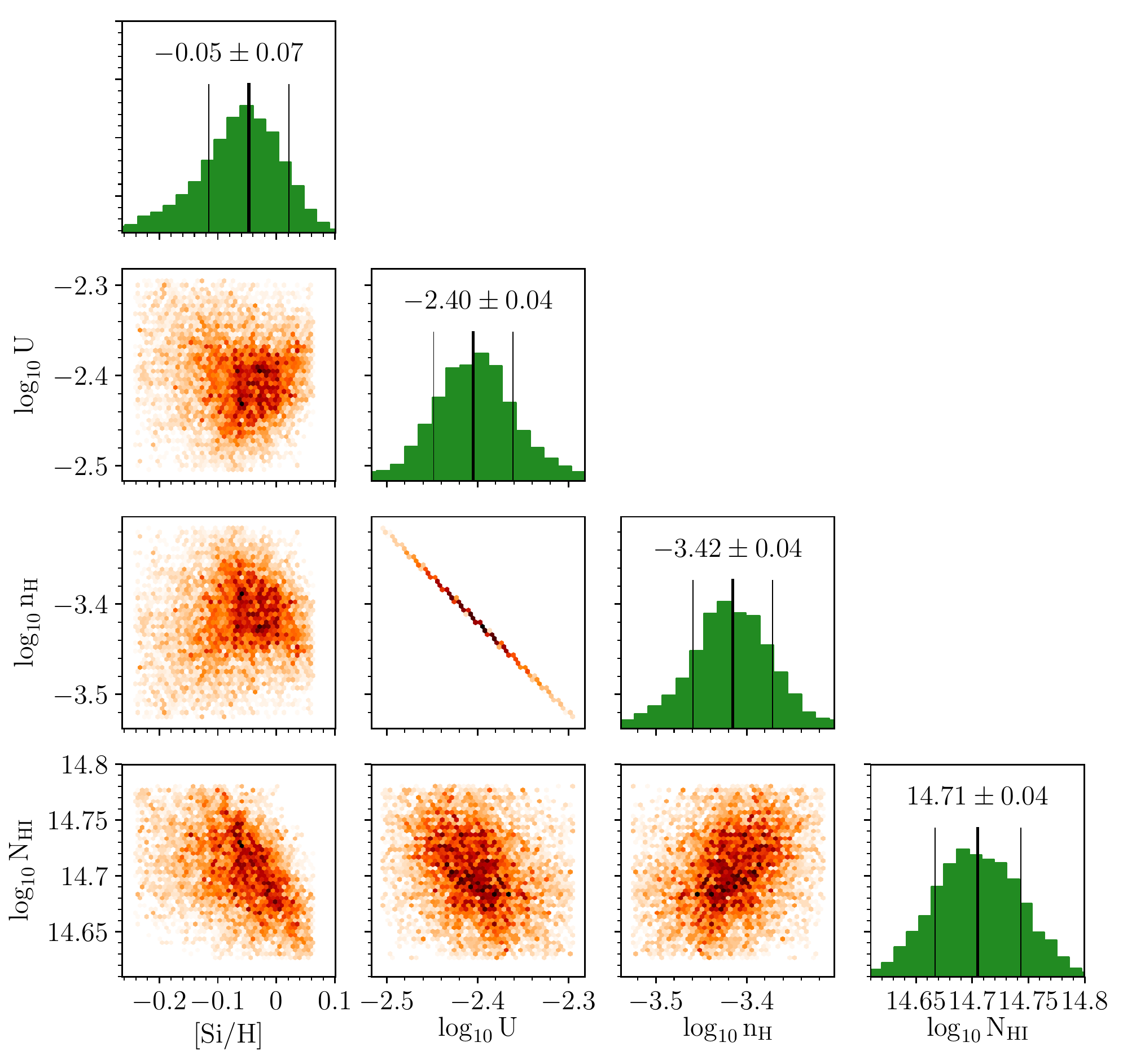}
\figsetgrpnote{The posterior distribution profiles from the MCMC analysis of the Cloudy grids for J$0853$, $z_{abs} = 0.0909$. The model parameters are plotted as described in Figure \ref{fig:Q0122_0.2119}.}
\figsetgrpend

\figsetgrpstart
\figsetgrpnum{2.6}
\figsetgrptitle{J$0910$, $z_{abs} = 0.2644$}
\figsetplot{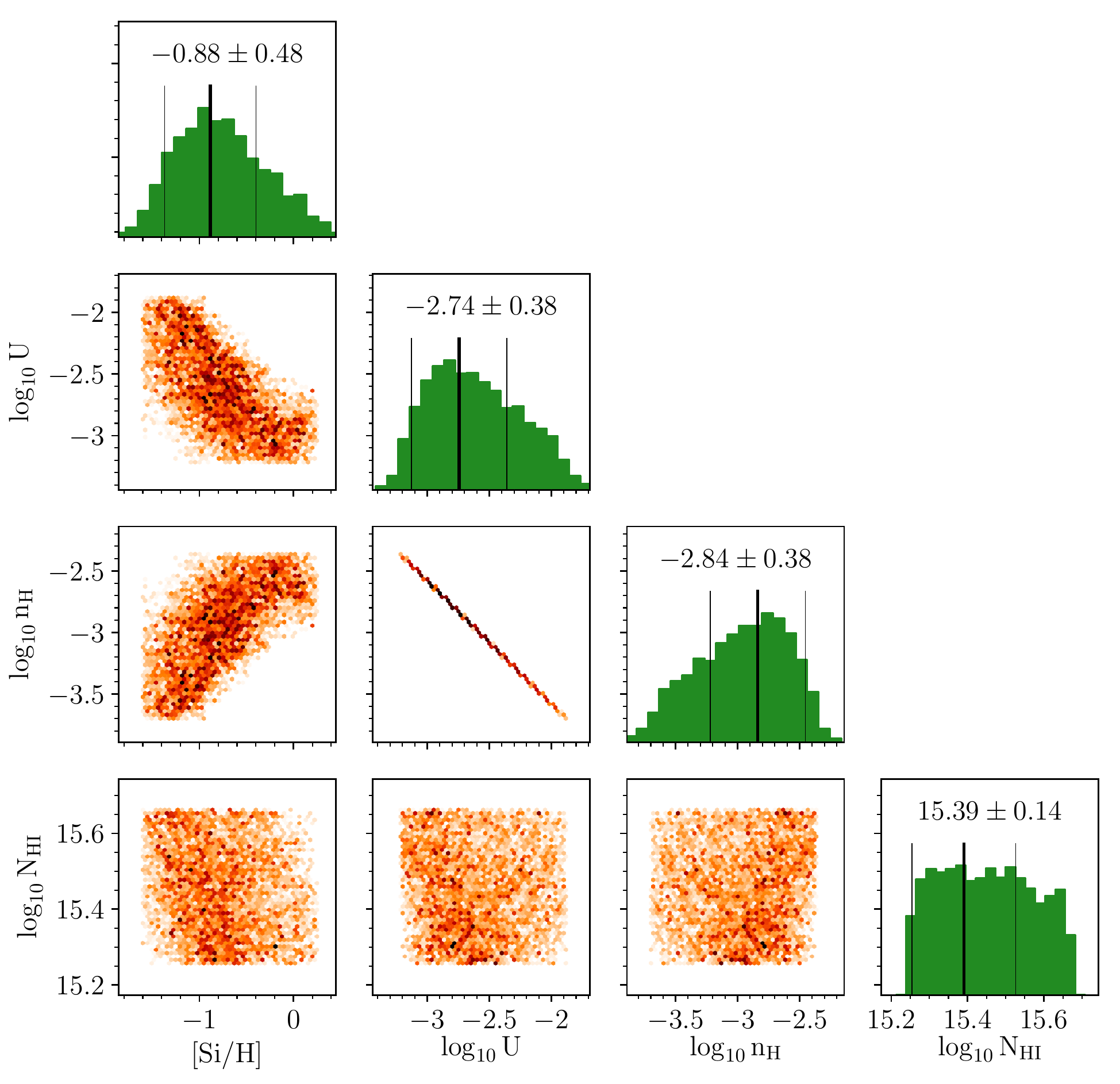}
\figsetgrpnote{The posterior distribution profiles from the MCMC analysis of the Cloudy grids for J$0910$, $z_{abs} = 0.2644$. The model parameters are plotted as described in Figure \ref{fig:Q0122_0.2119}.}
\figsetgrpend

\figsetgrpstart
\figsetgrpnum{2.7}
\figsetgrptitle{J$0925$, $z_{abs} = 0.2471$}
\figsetplot{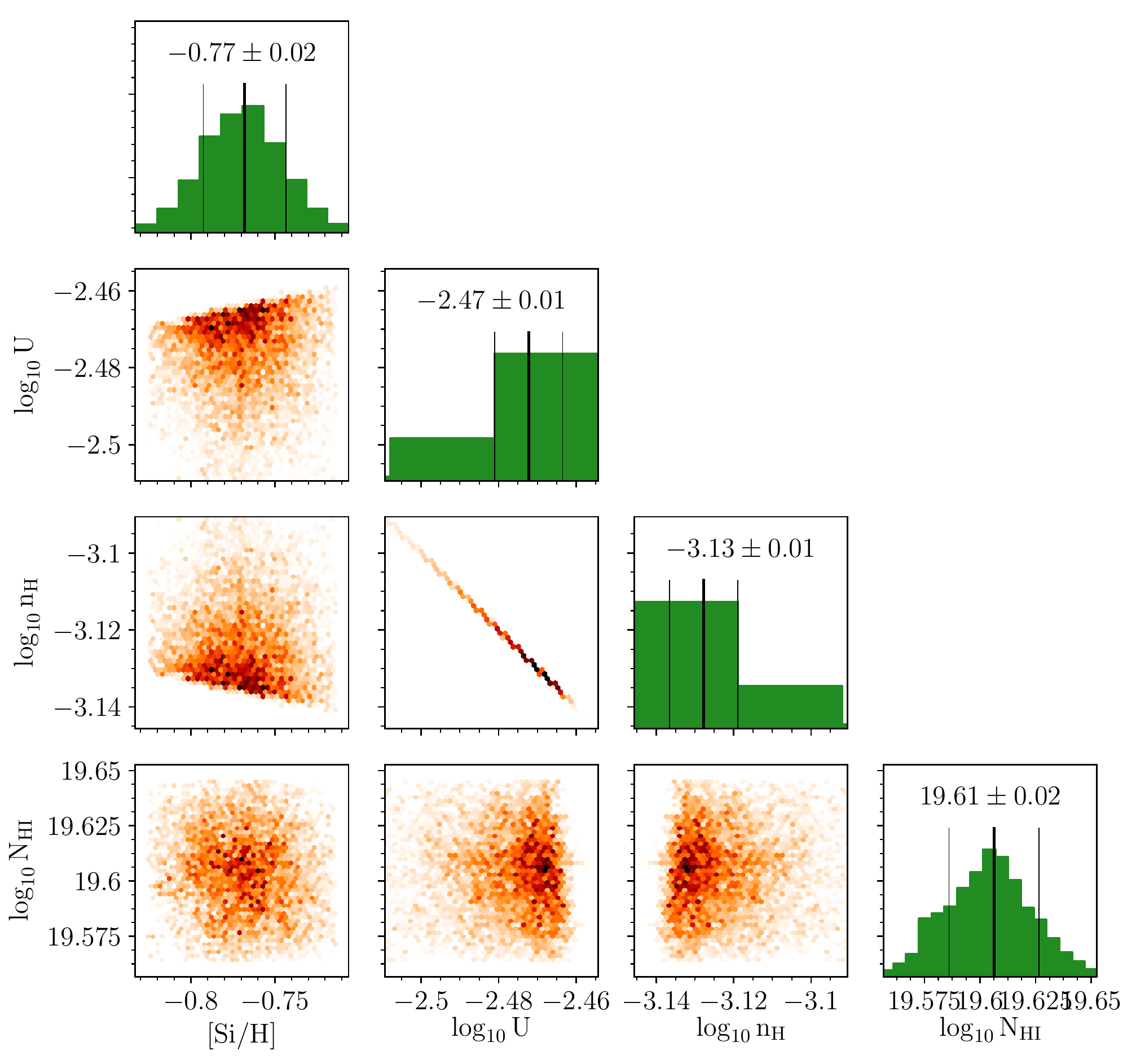}
\figsetgrpnote{The posterior distribution profiles from the MCMC analysis of the Cloudy grids for J$0925$, $z_{abs} = 0.2471$. The model parameters are plotted as described in Figure \ref{fig:Q0122_0.2119}.}
\figsetgrpend

\figsetgrpstart
\figsetgrpnum{2.8}
\figsetgrptitle{J$0928$, $z_{abs} = 0.1540$}
\figsetplot{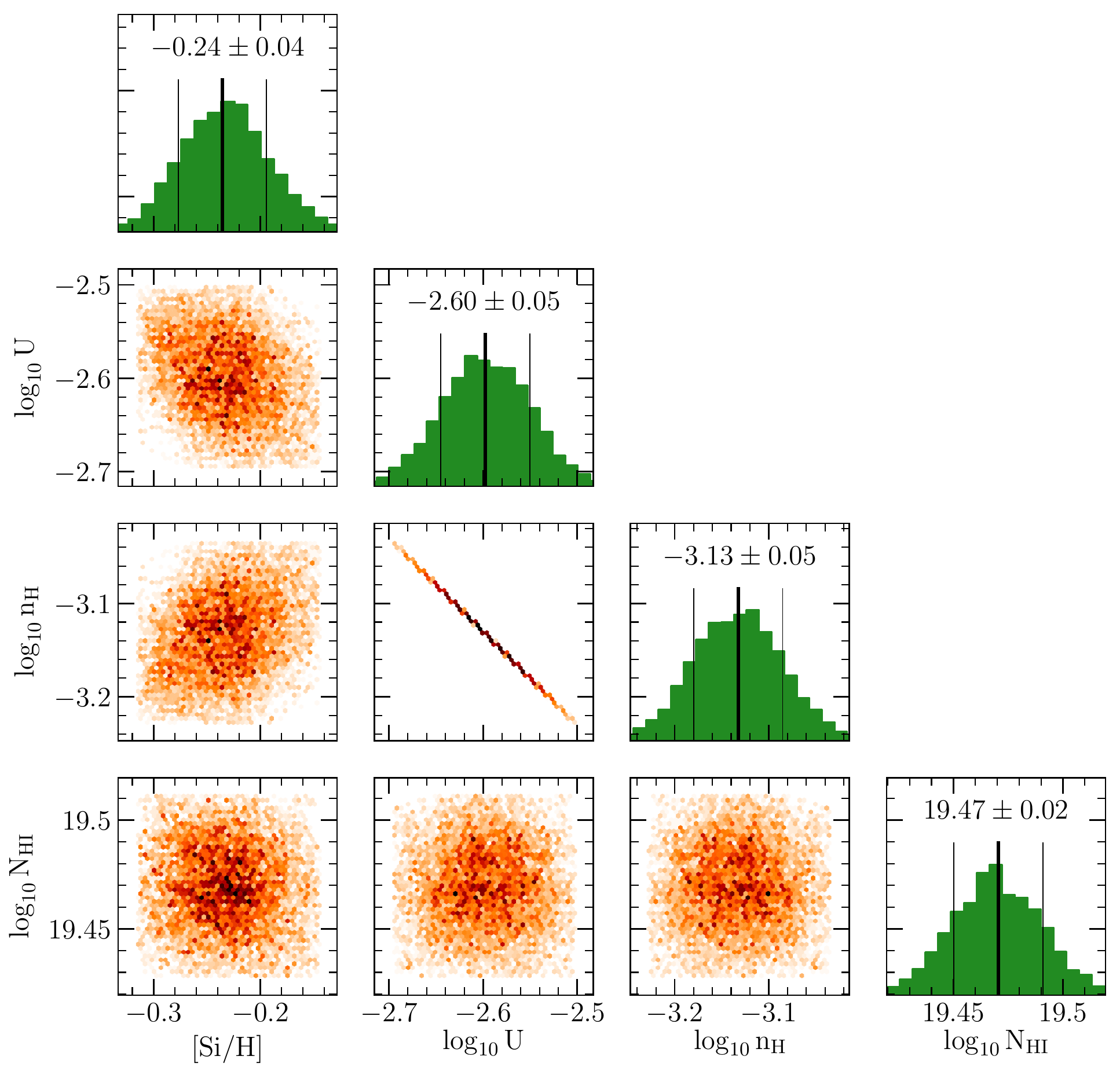}
\figsetgrpnote{The posterior distribution profiles from the MCMC analysis of the Cloudy grids for J$0928$, $z_{abs} = 0.1540$. The model parameters are plotted as described in Figure \ref{fig:Q0122_0.2119}.}
\figsetgrpend

\figsetgrpstart
\figsetgrpnum{2.9}
\figsetgrptitle{J$1009$, $z_{abs} = 0.3556$}
\figsetplot{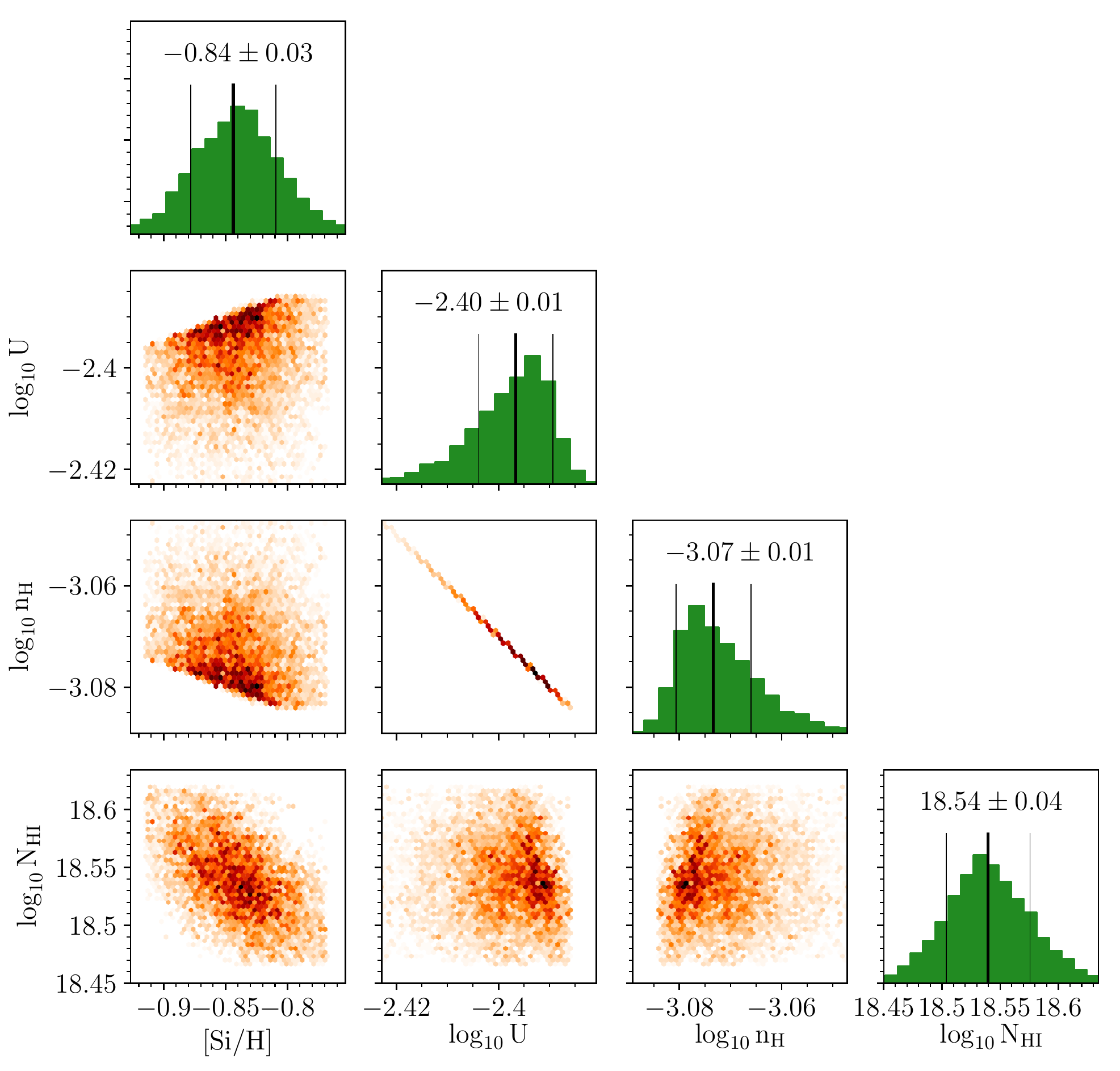}
\figsetgrpnote{The posterior distribution profiles from the MCMC analysis of the Cloudy grids for J$1009$, $z_{abs} = 0.3556$. The model parameters are plotted as described in Figure \ref{fig:Q0122_0.2119}.}
\figsetgrpend

\figsetgrpstart
\figsetgrpnum{2.10}
\figsetgrptitle{J$1119$, $z_{abs} = 0.0600$}
\figsetplot{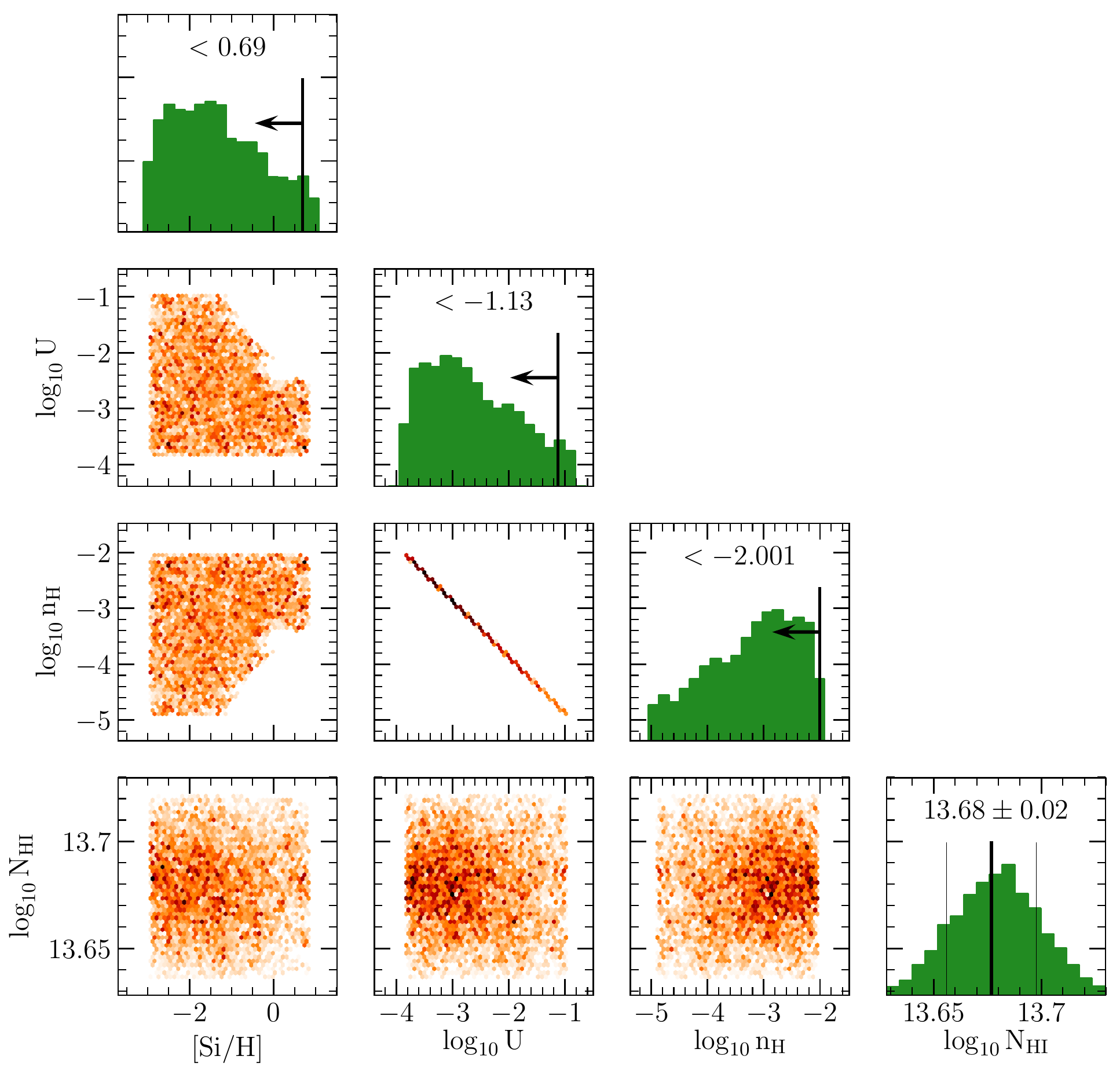}
\figsetgrpnote{The posterior distribution profiles from the MCMC analysis of the Cloudy grids for J$1119$, $z_{abs} = 0.0600$. The model parameters are plotted as described in Figure \ref{fig:Q0122_0.2119}.}
\figsetgrpend

\figsetgrpstart
\figsetgrpnum{2.11}
\figsetgrptitle{J$1133$, $z_{abs} = 0.2366$}
\figsetplot{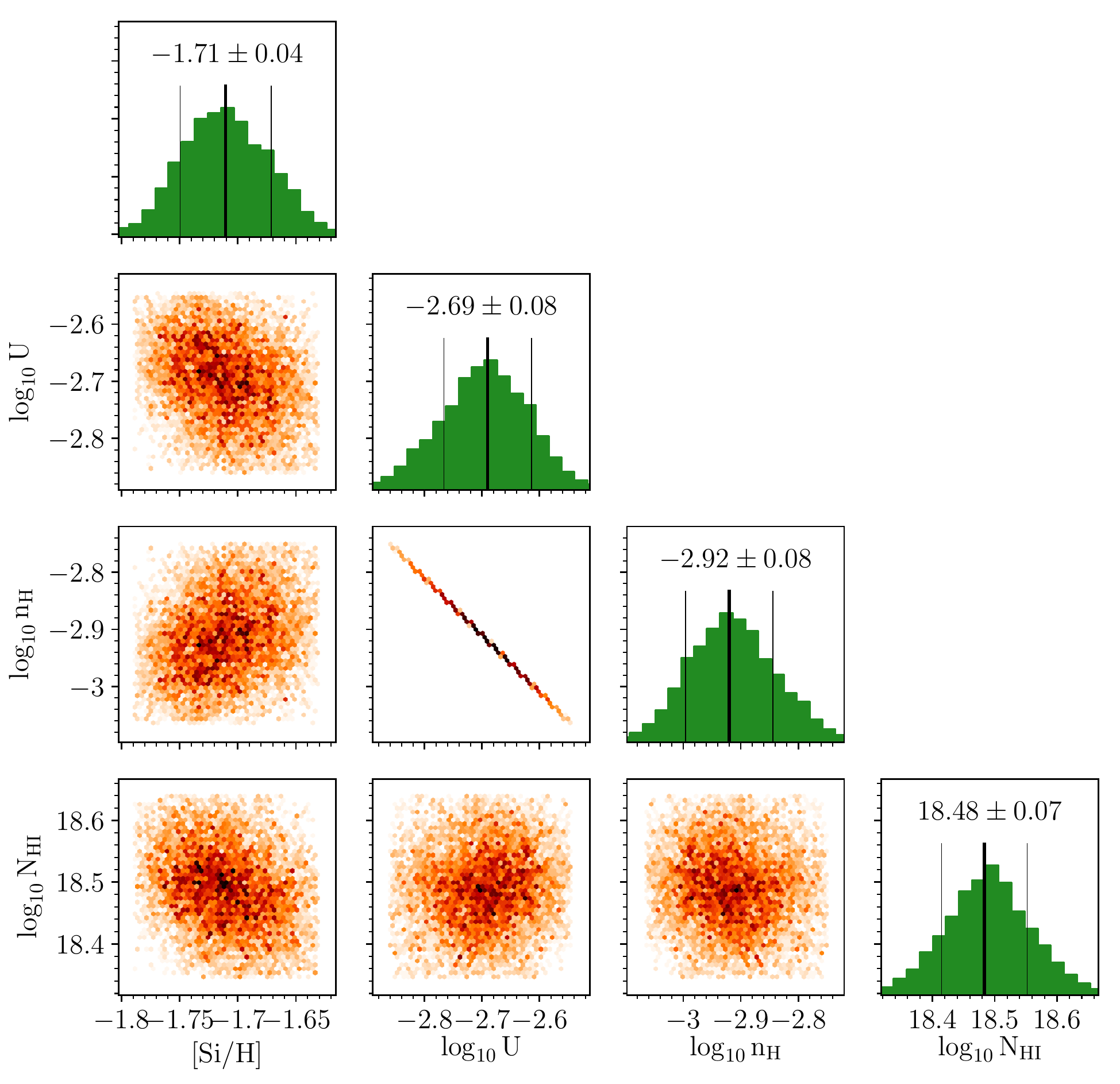}
\figsetgrpnote{The posterior distribution profiles from the MCMC analysis of the Cloudy grids for J$1133$, $z_{abs} = 0.2366$. The model parameters are plotted as described in Figure \ref{fig:Q0122_0.2119}.}
\figsetgrpend

\figsetend

\begin{figure}
	\centering
	\includegraphics[width=\linewidth]{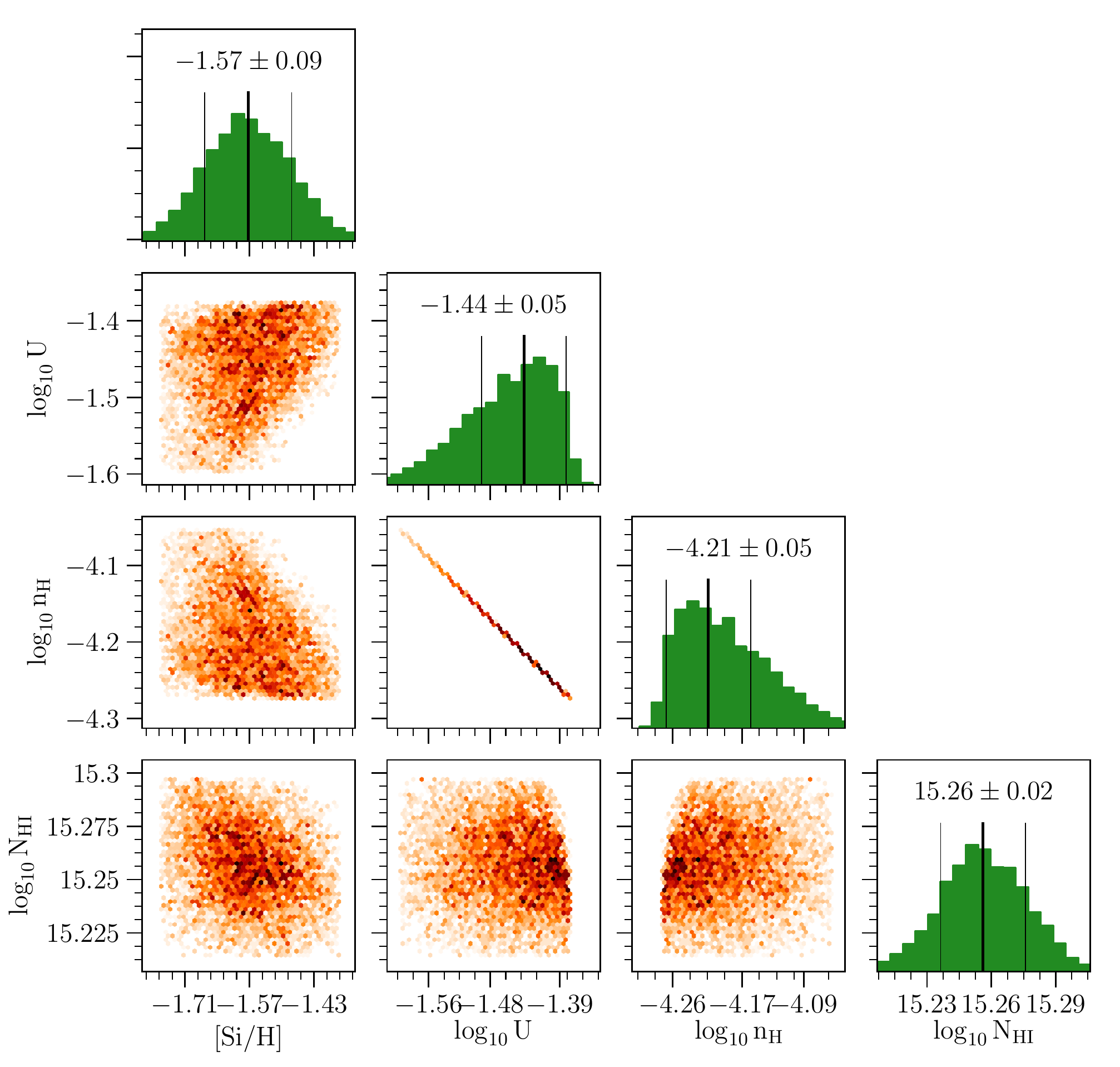}
	\caption{Posterior distribution profiles from the MCMC analysis of the Cloudy grids for J$0228$, $z_{abs} = 0.2073$ are shown as the orange hexbin plots. The model parameters shown are [Si/H], $\log U$, $\log n_H$ and ${\NHI}$. On the end of each row, the distributions of each of those parameters are shown in green where the $68\%$ confidence levels and the mean are shown above and indicated the black vertical lines. Plots for the rest of the sample are shown in Figure Set 2}
	\label{fig:Q0122_0.2119_par}
\end{figure}

The MCMC posterior distributions and histograms for J0228, $z_{abs} = 0.2073$ are shown in Figure \ref{fig:Q0122_0.2119_par}. The columns are plotted as a function of the metallicity, [Si/H], ionization parameter, $\log U$, hydrogen number density, $\log n_{H}$ and {\HI} column density, ${\NHI}$, from left to right. The posterior distributions of the MCMC walkers are shown. Darker orange indicates regions of higher probability. The final distributions of the MCMC walkers for each parameter are the green histograms at the end of each row, with the $68\%$ uncertainties and their average or the $95\%$ upper limit labelled above and indicated by black lines. The plots for the remaining systems are shown in Figure Set 2. Table \ref{tab:ionization_param} shows the inferred model parameters for the full sample, where we quote the most likely value, using the $68\%$ level as the uncertainty or the $95\%$ level for an upper limit.

\section{Results}
\label{sec:results}
Here we present the group environment CGM metallicities and their relation with HI column densities and impact parameters. We also compare the group environment properties to the isolated galaxy properties  presented in \citet{pointon19}.
 
\begin{figure*}
	\centering
	\includegraphics[width=\linewidth]{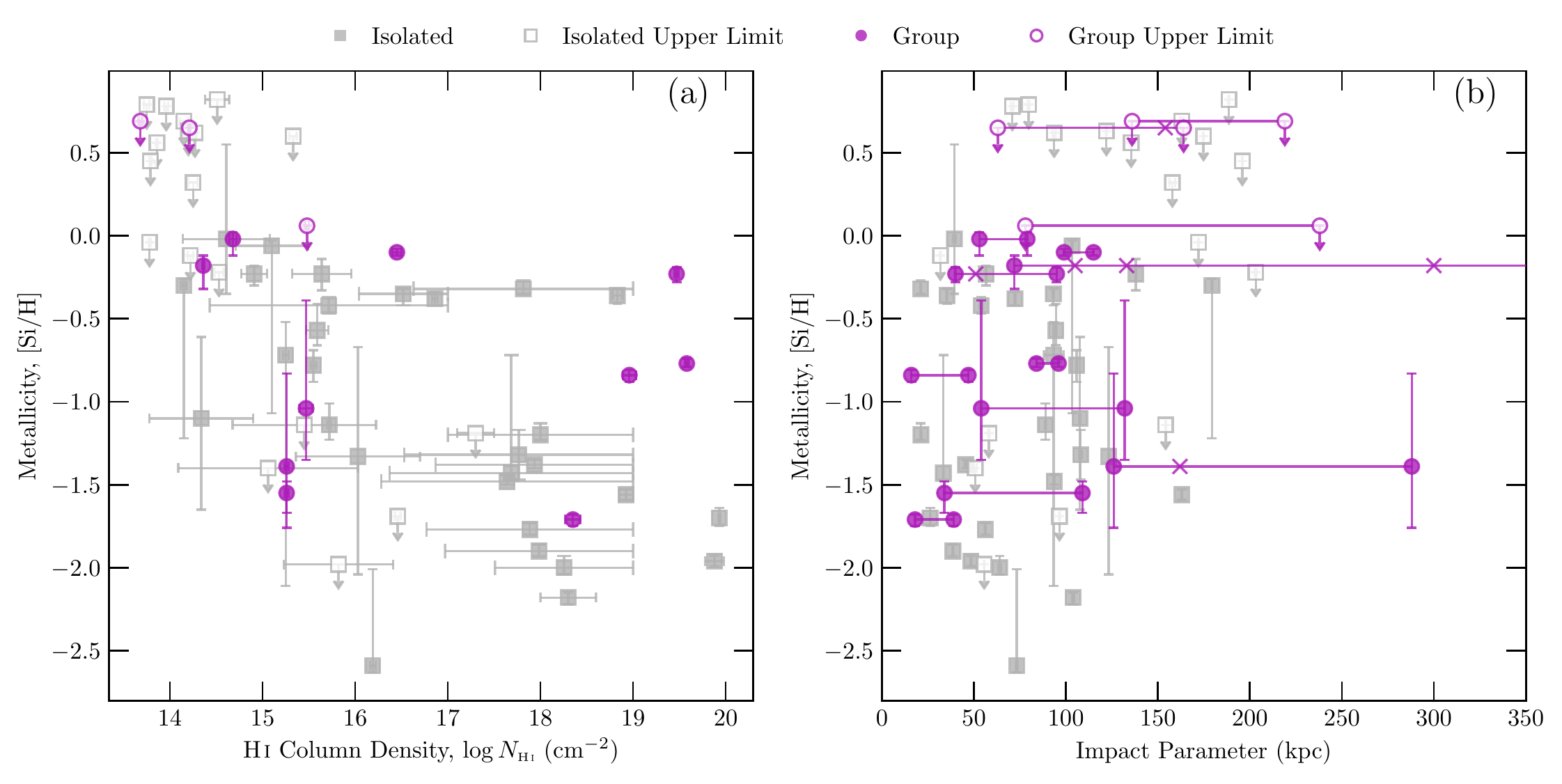}
	\caption{CGM metallicities for group and isolated environments as a function of (a) {\HI} column density and (b) impact parameter. Group environment CGM metallicities are purple filled circles, while metallicity upper limits are purple open circles. Isolated environment CGM metallicities are grey squares, while the metallicity upper limits are grey open squares. Since a group has, by definition, multiple galaxies associated with a given absorption system, we plot the impact parameter of each galaxy in a group and connect these galaxies with a horizontal line. Group galaxies that are nearest to and farthest from the quasar sightline are plotted as circles, while galaxies at intermediate impact parameters are plotted as crosses.}
	\label{fig:met_nh_d}
\end{figure*}

It is possible that group environments may alter the CGM metallicity. To test this, we compare the metallicity as a function of {\HI} column density between group and isolated environments in Figure \ref{fig:met_nh_d}(a). Group absorbers are purple circles while isolated absorbers are grey squares. Filled symbols indicate metallicity measurements, while open symbols represent limits. The group environment sample appears to overlap the isolated sample. A 1D Anderson-Darling test which accounts for upper limits indicated that the metallicity distribution of the group and isolated environment absorbers are drawn from the same population ($2.4\sigma$; $1.86\sigma$ for the $z<0.4$ isolated sample).  The median (mean) metallicity for group and isolated environments are [Si/H]~$=-1.04\pm0.29$ ($-1.07\pm0.23$) and [Si/H]~$=-1.20\pm0.16$ ($-1.25\pm0.13$), respectively. There is no significant difference between the median or mean metallicity for group environments compared to isolated galaxies ($0.4\sigma$ or $0.5\sigma$, respectively). Similarly, the median (mean) metallicity for the $z<0.4$ isolated environments is [Si/H]~$=-0.76\pm0.18$ ($-0.95\pm0.14$) and hence there is no significant difference compared to the isolated sample for the median or mean ($0.6\sigma$ or $0.3\sigma$, respectively).

Using Illustris simulations, \citet{hani18} studied the effect of a major merger on the CGM. They found that the post-merger CGM metallicity was $0.2$--$0.3$~dex higher than pre-merger. The difference in metallicity pre- and post-merger can be considered to be an upper limit on the expected metallicity difference between group and isolated environments since the galaxies in groups may not be in the post-merger phase. However, we note that the metallicity scatter in both the isolated galaxy and group environment samples is large ($>2$~dex), while the sample sizes are relatively small. This limits our ability to observe a metallicity difference of $0.3$~dex between the samples. 

To test our ability to detect a significant difference between our samples, we attempt to predict the number of group environments that are needed to detect a $3\sigma$ difference between the mean metallicity of the group and isolated environments. We use \begin{math} N = (3\times (\sigma_{g} + \sigma_{i}) / |\mu_{g} - \mu_{i}|)^2\end{math}, where the mean and error on the mean for the group environments are represented by $\mu_{g}$ and $\sigma_{g}$, respectively.  Similarly, the mean and error on the mean for the isolated environment sample is given by $\mu_{i}$ and $\sigma_{i}$, respectively. We determine that we would need to observe at least 36 group environments to observe a difference in the mean metallicity, assuming that the observed distributions are representative of the true metallicity distribution.

Previous studies have not detected an anti-correlation between the {\HI} column density and the CGM metallicity of isolated galaxies when the HM05 ionizing background was used in the Cloudy model \citep{chen17, zahedy18, wotta18, pointon19}. The presence of an anti-correlation in \citet{prochaska17} is thought to be due to the use of the HM12 ionizing background \citep{wotta18}. We also test for the presence of an anti-correlation between the {\HI} column density and CGM metallicity for group environments. A Kendall-tau rank correlation test, which accounts for metallicity upper limits, finds that we do not detect a significant anti-correlation between group environment CGM metallicity and {\HI} column density ($0.2\sigma$). The details of this test and additional Kendall-tau rank correlation tests are shown in Table \ref{tab:KTtest}. This is consistent with the non-detection of an anti-correlation between the CGM metallicity and {\HI} column density for isolated galaxies \citep[$2.1\sigma$;][]{pointon19}. Our ability to detect an anti-correlation is dependent on the size of the sample. Due to the large scatter and limited sample size of the metallicity in group environments, it is impossible to rule out the presence of an anti-correlation.

\begin{deluxetable*}{ccccccc}
	\tablecolumns{7}
	\tablewidth{0pt}
	\setlength{\tabcolsep}{0.06in}
	\tablecaption{Kendall-Tau Test Results\tablenotemark{a} \label{tab:KTtest}}
	\tablehead{
	    \colhead{Sample}   &
	    \colhead{Independent Variable}   &
		\colhead{Dependent Variable}           	&
        \colhead{Tau Statistic}     &
        \colhead{$p$-value}&
		\colhead{Confidence Level}               &
		\colhead{$\sigma$}  }
	\startdata
	\cutinhead{Rank Correlation Tests for the Isolated Sample without COS Halos galaxies (see Section 2.5)}
	Isolated	&	Closest Galaxy Impact Parameter, $D$	&	$\HI$ Column Density, $N_{HI}$	&$	0.00	$&$	<0.01	$&$	99.52	$&$	2.82	$	\\
Isolated, [Si/H]$ > -1.0$	&	Closest Galaxy Impact Parameter, $D$	&	$\HI$ Column Density, $N_{HI}$	&$	0.00	$&$	0.80	$&$	19.54	$&$	0.25	$	\\
Isolated, [Si/H]$ < -1.0$	&	Closest Galaxy Impact Parameter, $D$	&	$\HI$ Column Density, $N_{HI}$	&$	0.00	$&$	0.19	$&$	81.43	$&$	1.32	$	\\[-5pt]
	\cutinhead{Rank Correlation Tests for the Full Isolated Sample and the Group Sample}
	Group	&	$\HI$ Column Density, ({$\NHI$})	&	Metallicity, ([Si/H])	&$	-0.38	$&$	0.86	$&$	14.23	$&$	0.18	$	\\
Group	&	Closest Galaxy Impact Parameter, ($D$)	&	Metallicity, ([Si/H])	&$	-0.38	$&$	0.33	$&$	67.48	$&$	0.98	$	\\
Group	&	Closest Galaxy Impact Parameter, ($D$)	&	$\HI$ Column Density, ({$\NHI$})	&$	0.00	$&$	0.25	$&$	75.45	$&$	1.16	$	\\
Group, [Si/H]$ \geq -1.0$	&	Closest Galaxy Impact Parameter, ($D$)	&	$\HI$ Column Density, ({$\NHI$})	&$	-0.14	$&$	0.88	$&$	11.94	$&$	0.15	$	\\
Group, [Si/H]$ < -1.0$	&	Closest Galaxy Impact Parameter, ($D$)	&	$\HI$ Column Density, ({$\NHI$})	&$	0.00	$&$	0.22	$&$	77.93	$&$	1.22	$	\\
Isolated	&	Closest Galaxy Impact Parameter, ($D$)	&	$\HI$ Column Density, ({$\NHI$})	&$	2.89	$&$	<0.01	$&$	99.93	$&$	3.38	$	\\
Isolated, [Si/H]$ \geq -1.0$	&	Closest Galaxy Impact Parameter, ($D$)	&	$\HI$ Column Density, ({$\NHI$})	&$	0.00	$&$	0.34	$&$	66.30	$&$	0.96	$	\\
Isolated, [Si/H]$ < -1.0$	&	Closest Galaxy Impact Parameter, ($D$)	&	$\HI$ Column Density, ({$\NHI$})	&$	0.00	$&$	0.12	$&$	88.25	$&$	1.57	$\\[-5pt]	
	\cutinhead{Rank Correlation Tests for the Isolated Sample $z<0.4$}
	Isolated	&	Closest Galaxy Impact Parameter, $D$	&	$\HI$ Column Density, $N_{HI}$	&$	0.00	$&$	<0.01	$&$	99.23	$&$	2.67	$	\\
Isolated, [Si/H]$ > -1.0$	&	Closest Galaxy Impact Parameter, $D$	&	$\HI$ Column Density, $N_{HI}$	&$	0.00	$&$	0.34	$&$	66.30	$&$	0.96	$	\\
Isolated, [Si/H]$ < -1.0$	&	Closest Galaxy Impact Parameter, $D$	&	$\HI$ Column Density, $N_{HI}$	&$	0.00	$&$	0.24	$&$	75.51	$&$	1.16	$\\[-10pt]	
	\enddata
	\tablenotetext{a}{We use the Kendall-Tau formulation described by \citet{brown73} which accounts for upper limits, as implemented in ASURV \citep{feigelson85, feigelson86, feigelson90}.}
	
\end{deluxetable*}
In Figure \ref{fig:met_nh_d}(b), we present the CGM metallicity as a function of impact parameter for the group environment and isolated galaxy samples. The groups are purple, where the nearest and furthest galaxy members from the quasar sight-line are represented by circles, while any other group members are marked by a purple cross, all joined by a line. The isolated galaxy--absorber pairs are grey squares. Closed symbols represent metallicity measurements, while open symbols represent metallicity upper limits. We perform Anderson-Darling tests, which accounts for upper limits, comparing the impact parameter distributions of the isolated sample to three different measures of impact parameter in group environments: the nearest galaxy member, the mean impact parameter and the most luminous galaxy. We find that the differences between the isolated galaxy and the three group environment impact parameter distributions are statistically insignificant ($1.7\sigma$, $0.4\sigma$ and $0.2\sigma$, respectively). Similarly, the difference between the $z<0.4$ isolated environment sample and the three group environment impact parameter distributions are statistically insignificant ($1.8\sigma$, $0.1\sigma$ and $0.1\sigma$) Additionally we test for a correlation between the group CGM metallicity and the nearest galaxy impact parameter by doing a Kendall-tau rank correlation test, taking upper limits into account. We do not detect a significant relationship ($1.0\sigma$). This is consistent with \citet{pointon19} who found no trend between the impact parameter of isolated galaxies and the CGM metallicity, although metallicities are rarely measured beyond $120$~kpc due to a lack of metal detections.

\begin{figure*}
	\centering
	\includegraphics[width=\linewidth]{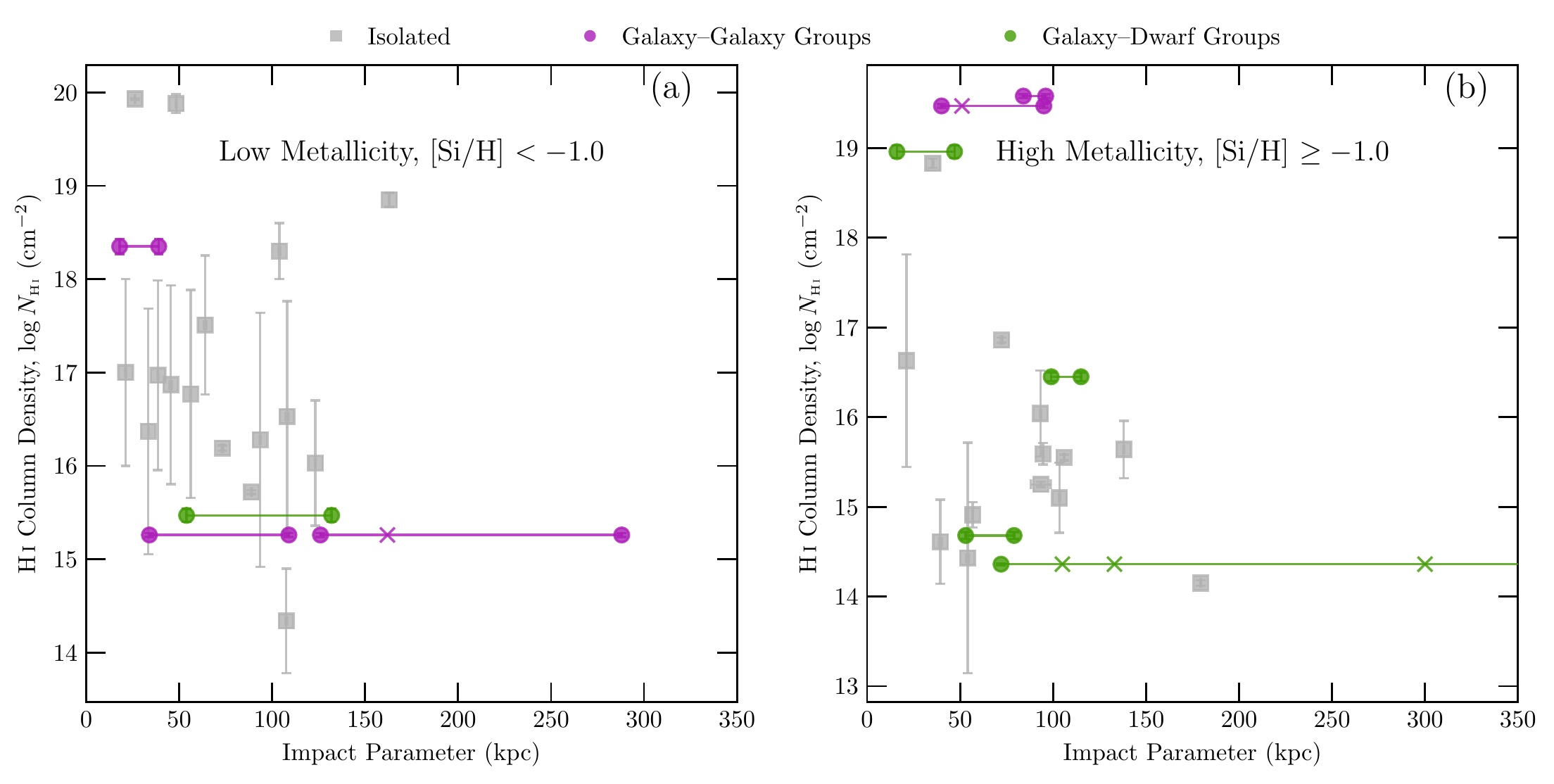}
	\caption{{\HI} column densities of group and isolated environments as a function of impact parameter for (a) low metallicity ([Si/H]$<-1.0$) and (b) high metallicity ([Si/H]$\geq -1.0$). Galaxy-galaxy and galaxy-dwarf group environments are in purple or green, respectively. For a given group environment, galaxies nearest to and farthest from the quasar sightline are plotted as circles, while the rest of the group member galaxies are plotted as crosses. A horizontal line is plotted for each group environment to indicate group membership. Isolated galaxies are plotted as gray squares. Limits for both the isolated and group sample have been removed since it is impossible to determine if they correspond to high or low metallicity gas. }
	\label{fig:nh_d_metcomp}
\end{figure*}

\citet{lehner18review} presented results investigating the metallicity of 6 pLLS+LLS where they identified more than one potential galaxy which could be associated with the absorption feature. Due to the preliminary nature of the results, the authors refrain from drawing any conclusions from the data. However, they suggest it may be possible that for [X/H]$\geq -1$ the absorption systems are more likely to be associated with group environments. In contrast, absorption systems with [X/H]$<-1$, may be associated with individual galaxies in all but one absorber. We investigate this possible cause of a metallicity bimodality due to group environments by bifurcating the group and isolated samples at [Si/H]$=-1$ and comparing the relationship between the {\HI} column density and impact parameter for high and low metallicities.

In both panels of Figure \ref{fig:nh_d_metcomp}, the {\HI} column density is plotted as a function of metallicity. Group environments are purple or green circles, while isolated galaxies are grey squares. Figure \ref{fig:nh_d_metcomp}(a) shows [Si/H]$<-1$ while Figure \ref{fig:nh_d_metcomp}(b) shows [Si/H]$\geq -1$. We exclude upper limits from the plot and analysis since it is impossible to determine if they refer to high or low metallicity gas. While preliminary results from \citet{lehner18review} found that 6/7 group environments had high metallicity CGM, we find 6/10 with high metallicity. We note that we probe a disparate {\HI} column density range compared to \citet{lehner18review} who investigated pLLS and LLS with a {\HI} column density range of $16.2 < {\NHI} < 19.0$. However, since a comparable range of metallicities and {\HI} column densities are observed for both group environments surveyed in this paper and isolated galaxies in \citet{pointon19}, we suggest that the peaks of the metallicity bimodality observed by \citet{lehner13} and \citet{wotta16, wotta18} are not driven by environment.

Additionally, we test for an anti-correlation between the {\HI} column density and the impact parameter for group environments using a Kendall-tau rank correlation test for the entire sample ($1.2\sigma$), low metallicity absorbers ($1.2\sigma$) or high metallicity absorbers ($0.2\sigma$), where the nearest galaxy impact parameter was used. For the isolated sample, a Kendall-tau rank correlation test finds that the entire sample has a significant anti-correlation between the {\HI} column density and the impact parameter ($3.9\sigma$). The same test on the $z<0.4$ isolated environment sample does not find a significant anti-correlation between the {\HI} column density and the impact parameter ($2.7\sigma$), due to the smaller sample size. Isolated galaxy high and low metallicity absorbers do not have significant anti-correlations between the {\HI} column density and impact parameter ($1.0\sigma$ and $1.6\sigma$, respectively for the full sample; $1.0\sigma$ and $1.2\sigma$, respectively for the $z<0.4$ sample), likely due to the lower number of systems in each bin. Similarly, the lack of anti-correlations between the {\HI} column density and nearest galaxy impact parameter for the entire group sample is most likely due to the limited sample size in group environments. However, if the lack of anti-correlation is due to a physical process in group environments such as tidal stripping, higher metallicity gas may be distributed to larger impact parameters.

It is plausible to expect that the CGM of galaxies with similar mass may be affected differently to those with higher mass ratios. \citet{nielsen18} found that groups with similar mass galaxies may have larger {\MgII} equivalent widths and velocity dispersions compared to groups with differing mass galaxies. Assuming that the $B$-band luminosity is a proxy for galaxy mass, we define galaxy--galaxy groups, which may later form a major merger, as those with a luminosity ratio between the most and second most luminous galaxies of $L_1/L_2<3.0$. In contrast, a galaxy--dwarf group, which may become a minor merger, has a luminosity ratio of $L_1/L_2 \geq 3.0$. To probe the effect of mass ratios on the CGM metallicity we compare high and low metallicity for galaxy--galaxy and galaxy--dwarf groups in Figure \ref{fig:nh_d_metcomp}. Galaxy--galaxy groups are purple circles, while galaxy--dwarfs are green circles. 

We find that all but one galaxy--dwarf groups have high metallicities, while galaxy--galaxy groups have both high and low metallicities. The metallicity medians (means) for the galaxy--galaxy and galaxy--dwarf group environments are $-0.8\pm0.4$ ($-0.7\pm0.3$) and $-0.2\pm0.3$ ($-0.2\pm0.1$), respectively. The median (mean) metallicities for the galaxy--galaxy and galaxy--dwarf samples differ by $1.9\sigma$ ($1.1\sigma$) and thus it is unclear if the masses of the galaxies within group environments play a role in the enrichment of the CGM. 

The errors on the median or mean metallicities are highly dependent on the sample size, limiting our ability to find a significant difference in the metallicities of the galaxy--galaxy and galaxy--dwarf group environments. Therefore, we attempted to predict how many groups would be required to measure a $3\sigma$ difference between the two samples using \begin{math} N = (3\times (\sigma_{gd} + \sigma_{gg}) / |\mu_{gd} - \mu_{gg}|)^2\end{math}, where the mean and error on the mean for the galaxy--galaxy sample are represented by $\mu_{gg}$ and $\sigma_{gg}$, respectively. Similarly, the mean and error on the mean for the galaxy--dwarf sample is given by $\mu_{gd}$ and $\sigma_{gd}$, respectively. We determine that we would need to observe at least $6$ group environments in each subsample to detect a significant difference in the mean metallicity, assuming the observed distributions are representative of the true metallicity distribution. 

\section{Discussion}
\label{sec:discussion}
Our ``Multiphase Galaxy Halos'' Survey has probed the CGM metallicity in 13 $z < 0.4$ group environments over a range of {\HI} column densities ($13.6 < {\NHI} < 19.6$). Typical group environments have two members, although we detect up to five in J$0407$, $z = 0.0914$. We do not detect a relationship between {\HI} column density and metallicity in group environments ($0.2\sigma$), consistent with isolated environments \citep[$2.1\sigma$;][]{pointon19}. However, we note that the small sample size of the groups makes it difficult to investigate this further. The lack of anti-correlation is consistent with \citet{wotta18} who find that LLS and pLLS have a metallicity range of $-3<$[X/H]$<0$, which narrows to $-1.8<$[X/H]$<0$ for sub-DLAs and DLAs.

We not not detect an anti-correlation between the CGM metallicity and the impact parameter of the nearest group galaxy member. This is consistent with isolated environments where the presence of an anti-correlation could not be confirmed \citep{pointon19}, although simulations suggest an anti-correlation should be present \citep{crain13} since it is expected that gas metallicities should decrease to that of the IGM at larger impact parameters. Small number statistics and large scatter in the CGM metallicity of group environments may explain the non-detection of an anti-correlation with impact parameter. However, it may also be possible that the IGM at the low redshifts probed by \citet{pointon19} and this study is sufficiently polluted by outflows that the difference between the IGM and CGM metallicities has become difficult to detect with current sample sizes. Group environments further complicate the picture. \citet{nielsen18} found group environments can distribute cool {\MgII} gas through an intragroup medium. If it is assumed that {\MgII} detections are analogous to CGM metallicity detections, we can predict that an intragroup medium would also result in a flatter relationship between CGM metallicity and impact parameter. Unfortunately, it is not possible to determine if the lack of relationship between CGM metallicity and impact parameter in group environments results from an intragroup medium due to the small sample size.

It has been found in simulations that the post-merger CGM metallicity is $0.2$--$0.3$~dex higher than pre-merger \citep{hani18}. While the difference between the mean CGM metallicities of group and isolated environments is on on the order of $0.2$~dex, we do not detect a significant difference between the metallicity distributions. This indicates that CGM metallicity of group environments does not differ from isolated environments, although we predict that increasing the sample size of group environments to 36 may result in a significant difference. However, it is important to note that the group environments in this survey are loose groups and may not yet be gravitationally bound. The lack of difference between the CGM metallicity of group and isolated environments is consistent with the possibility that any interactions in the group environments have not yet had sufficient time to increase the metallicity and that differences between group and isolated environment metallicities may only be detected for major mergers similar to the event simulated by \citet{hani18}.

Using the FIRE simulations, \citet{angles16} found that the dominant accretion mechanism for CGM gas at $z<1$ was through intergalactic transfer, which \citet{pointon19} suggests could drive the large scatter found in the metallicity distribution of isolated environments. The presence of well-mixed CGM halos in isolated environments at $z<1$ is consistent with the view that CGM gas has already been enriched by mergers. \citet{hani18} found that the time required for the CGM metallicity to return to pre-merger values was at least $1$~Gyr. Based on the difference between the minimum and maximum galaxy redshifts and standard cosmology, the group environment sample covers a time span of $3.6$~Gyr while the isolated sample encompasses $5.3$~Gyr \citep{wright06}. Thus it is reasonable to expect that the CGM metallicities of galaxies which experienced a merger interactions over $1$~Gyr earlier may have reverted to their pre-merger values. Alternatively, since we cannot rule out past merger events for any isolated or group galaxy, it may be possible for the CGM in both environments to have been enriched such that there is no detectable difference in metallicity at low redshifts. It is likely that a combination of these two effects have resulted in the large CGM metallicity spreads found in low redshift isolated and group environments. 

Studies of the CGM have established that the {\HI} column density around isolated galaxies decreases with increasing impact parameter \citep[e.g.,][]{lanzetta95, tripp98, chen01a, rao11, borthakur15, curran16, prochaska17, pointon19}. However, given the flattened relationship between {\MgII} equivalent width and impact parameter for group environments \citep[e.g.][]{chen10a,bordoloi11,nielsen18}, it is reasonable to expect that the {\HI} may be similarly affected. Indeed, the {\HI} column density has no significant anti-correlation with the impact parameter of the nearest group galaxy. We also do not find any anti-correlation between the {\HI} column density and nearest galaxy impact parameter for either high ([Si/H]$\geq -1.0$) or low ([Si/H]$< -1.0$) metallicities. 

However, if {\MgII} is assumed to be a proxy for metallicity, the stronger equivalent widths in group environments found by \citet{chen10a, bordoloi11, nielsen18} are somewhat in tension with our finding that the CGM metallicity distribution is not significantly different to that of isolated galaxies. However, it is important to note that of the 7/13 group environments in the survey where {\MgII} was covered, there were 4 absorbers and 3 non-absorbers. The inferred metallicities for 9/13 galaxies were dependent on the high ionization states of {\NV}, {\SiIV} or {\CIV}. Therefore, the assumption that {\MgII} is a proxy for metallicity does not hold for all group environments. Studies of the highly ionized gas in group environments have found that oxygen and carbon tend to be ionized above {\OVI} and {\CIV} \citep[e.g][]{burchett16, oppenheimer16, pointon17, ng19}. Therefore, it is possible that combining a single phase gas model with highly ionized gas has lead to the metal content of carbon, nitrogen or silicon to be underestimated, resulting in lower metallicities. 

\citet{lehner18review} also investigated the relationship between {\HI} column density and impact parameter for group environments. While they did not draw any conclusions from their preliminary results they find a hint that low metallicity systems could be associated with isolated environments, while high metallicity absorbers may be associated with groups. Following this suggestion, after bifurcating the metallicity of group environments at centre of the bimodal metallicity distribution found by \citet{lehner13} and \citet{wotta16, wotta18} ([Si/H]$=-1.0$), we find that groups are scattered across both high (6/10) and low metallicities (4/10). This indicates that a wide range of CGM metallicities, as opposed to only high metallicities are associated with group environment and that galaxy environment may not be the source of the metallicity bimodality.

The mass ratios of the galaxies in group environments have been found to have an effect on the {\MgII} absorption \citep{nielsen18}, where groups with similar member galaxy luminosities may have larger equivalent widths and velocity dispersions than groups with different member galaxy luminosities. The authors suggested that galaxy--galaxy group environments may be more efficient at causing enhanced star formation and/or tidal stripping of gas.
We used the $B$-band luminosity ratio as a proxy for galaxy mass to classify groups as galaxy--galaxy group ($L_1/L_2 <3.0$) or galaxy--dwarf group ($L_1/L_2\geq3.0$) environments. Using the same metallicity cut ([Si/H]$=-1.0$) as \citet{lehner18review}, we found that all but one galaxy--dwarf group environments were associated with high metallicity gas, while galaxy--galaxy groups were associated with both high and low metallicity gas. Although, we did not find a significant difference in the median or mean metallicities of galaxy--galaxy group and galaxy--dwarf group environments, a simple model of the group subsamples predicts that observing nine or more high and low metallicity group environment CGM absorbers could find a significant difference between the means of the subsample metallicities. We note that many of the inferred metallicities rely on highly ionized gas phases. However, previous studies of {\OVI} and {\CIV} in group environments have found that the high halo mass is sufficient to push oxygen to higher ionization states \citep{burchett16, oppenheimer16, pointon17, ng19}. Galaxy-galaxy group environments may have sufficient mass to have further ionized carbon, nitrogen or silicon, resulting in an underestimation of the metal content of the gas, which in turn could result in lower metallicities. However, galaxy-dwarf environments may not yet have sufficient mass, and hence temperature, to further ionize the CGM.

The metallicities in this study were calculated using the total column density along the line-of-sight in the absorber. However, studies of the CGM metallicity around isolated environments have found that the metallicity is not constant across an absorption profile \citep[e.g.][]{churchill12, crighton15, muzahid15, rosenwasser18, zahedy18, peeples18}. In essence, this means that low metallicity gas along the line-of-sight can be obscured by the presence of high metallicity gas structures, such as accretion, outflows and tidal streams. This effectively masks any information about the structure of metals in the CGM. Future studies should attempt to calculate the metallicity structure of each individual absorber, which may assist in determining the characteristics of the CGM in group environments.

Although integrated line-of-sight metallicities may obscure some low metallicity gas, nearly half of the metallicity measurements in group environments are metal-poor. This clearly indicates there is a mix of high and low metallicity gas within group environments, similar to what is found in isolated galaxies. Using integrated line-of-sight metallicity values, it is difficult to determine if the metallicity is associated with an intragroup medium or individual galaxies within the group. Given that both {\OVI} or {\MgII} absorption are associated with an intragroup medium, rather than a superposition of individual halos, it is expected that other gas traces of the halo gas in group environments would have a similar structure \citep{pointon17, nielsen18}. Component-by-component metallicity studies could reveal how individual clumps of gas within the CGM halo are amalgamated into an intra-group medium.

\section{Summary and Conclusions}
\label{sec:conclusions}
We used the ``Multiphase Galaxy Halos'' Survey to calculate the CGM metallicity of $13$~$z<0.4$ group environments. The column density for each covered ion in the absorption systems was calculated using VPFIT. These column densities were compared to the predicted column densities from Cloudy ionization models using a MCMC analysis to infer a metallicity of the absorption systems. The CGM metallicity was then compared to the {\HI} column density of the CGM gas and the impact parameters of the group members. Our findings are:

\begin{enumerate}
    \item Group environment CGM metallicities span a large range of $-2<$[Si/H]$<0$ with a mean of $\langle$[Si/H]$\rangle=-0.54\pm0.22$. These are consistent with isolated galaxy CGM metallicities ($-3.0<$[Si/H]$<0$, $\langle$[Si/H]$\rangle=-0.77\pm0.14$) at the $0.6\sigma$ level. There is no significant enrichment of the group environment CGM at $z<0.4$. Indeed, the similar span of metallicities in group and isolated environments suggests that there is no general preferential association of group environments with high metallicity gas.
    \item We do not detect a significant anti-correlation between the CGM metallicity and the {\HI} column density ($0.2\sigma$) in group environments. This is consistent with previous studies which have used the HM05 ionizing background to infer CGM gas metallicities. 
    \item There is no significant anti-correlation between the metallicity and impact parameter of the nearest group galaxy, the mean impact parameter or the most luminous galaxy ($1.3\sigma$, $0.2\sigma$ and $0.1\sigma$, respectively). This is consistent with the absence of a relationship between the metallicity and impact parameter in isolated environments. It may be possible that at low redshifts, previous interactions have enriched the surrounding IGM, resulting in a lack of correlation between impact parameter and CGM metallicity.
    \item We do not detect a significant anti-correlation between the {\HI} column density and the impact parameter of the nearest galaxy in group environments. This is contrary to what is detected in the entire isolated sample where the {\HI} column density has been measured to decrease as the distance from the galaxy increases. Although the lack of anti-correlation in group environments may be due to low number statistics, the flattened relationship is consistent with {\MgII} and {\OVI} studies, which have found evidence for an intragroup medium.
    \item We further examine the environments of the groups by bifurcating the sample at $L_1/L_2=3.0$ and found median metallicities of $-0.8\pm0.4$  and $-0.2\pm0.3$ for low (galaxy--galaxy) and high (galaxy--dwarf) luminosity ratios, respectively. Although there is no significant difference between the median ($1.9\sigma$), all but one galaxy-dwarf metallicity measurements have [Si/H]$>-1.0$, while galaxy--galaxy group environments have both low and high metallicities. Larger samples should be able to determine if there is a difference between the CGM metallicities of galaxy--galaxy and galaxy--dwarf environments. 
\end{enumerate}
With our sample size, we are unable to confidently detect a significant enhancement in the CGM metallicity for group environments. While we do not find any metallicity enhancement here with environment, samples larger than 36 group environments may find a more metal-rich intragroup medium. Larger samples may also find that a large luminosity ratio, and hence mass ratio, of the galaxies involved increase the metallicity. Regardless, we expect that a strong, detectable metallicity enhancement may only occur when galaxies are in the process of interacting or merging, which is not represented in our sample. Future work should focus on creating samples of galaxy groups that are undergoing different phases of evolution, e.g, loose groups, compact groups, interactions and mergers, to fully understand how galaxy environment affects the evolution of the CGM metallicity. Furthermore, studies should focus on understanding how the CGM metallicity differs along the line-of-sight of each absorption profile so that high metallicity gas does not obscure metal-poor material. Such studies may also be able to use the information on the contribution of each group galaxy to the CGM to test for an intragroup medium or a superposition model using metallicity as a tracer.

\acknowledgments

We would like to thank John O'Meara for providing HIRES spectra, B. Wakker for providing the \textit{FUSE} spectra, Nicolas Lehner for discussions on the UV ionizing background, and Neil Crighton for the MCMC analysis software and Cloudy ionization training. 
Support for this research was provided by NASA through grants \textit{HST} GO-13398 from the Space Telescope Science Institute, which is operated by the Association of Universities for Research in Astronomy, Inc., under NASA contract NAS5-26555. 
S.K.P acknowledges support through the Australian Government Research Training Program Scholarship. 
G.G.K, N.M.N, and M.T.M acknowledge the support of the Australian Research Council through the Discovery Project DP170103470. 
Parts of this research were supported by the Australian Research Council Centre of Excellence for All Sky Astrophysics in 3 Dimensions (ASTRO 3D), through project number CE170100012.
Some of the data presented herein were obtained at the W. M. Keck Observatory, which is operated as a scientific partnership among the California Institute of Technology, the University of California and the National Aeronautics and Space Administration. 
Observations were supported by Swinburne Keck programs 2014A\_W178E, 2014B\_W018E, 2015\_W018E, 2016A\_W056E and 2017A\_W248. 
The Observatory was made possible by the generous financial support of the W. M. Keck Foundation.  
The authors wish to recognize and acknowledge the very significant cultural role and reverence that the summit of Maunakea has always had within the indigenous Hawaiian community.  
We are most fortunate to have the opportunity to conduct observations from this mountain.  
Based on observations collected at the European Organisation for Astronomical Research in the Southern Hemisphere under ESO programs listed in Table \ref{tab:obsgal}.

\bibliographystyle{apj}
\bibliography{refs}

\end{document}